\documentclass[preprint,3p,12pt, authoryear]{elsarticle}
\journal{XX SUBMITTED TO JOURNAL - UNDER REVIEW}
\makeatletter
\def\ps@pprintTitle{%
	\let\@oddhead\@empty
	\let\@evenhead\@empty
	\def\@oddfoot{}%
	\let\@evenfoot\@oddfoot}
\makeatother

\usepackage{geometry}
\usepackage[german,english]{babel}
\usepackage[utf8]{inputenc}
\usepackage[T1]{fontenc}
\usepackage[german=guillemets,autostyle = true]{csquotes}
\MakeAutoQuote{«}{»}
\usepackage{graphicx}
\usepackage{amsmath}
\usepackage{amssymb}
\usepackage{amsfonts}
\usepackage{amsthm}
\usepackage{bm}
\usepackage{arydshln}
\usepackage{bbm} 
\usepackage{mathrsfs}
\usepackage[mathcal]{euscript}
\usepackage{mathtools}
\usepackage{mathptmx}
\usepackage{verbatim}
\usepackage{enumitem}

\usepackage{lipsum} % Nur fuer den Blindtext
\usepackage[ruled]{algorithm2e}
\SetKwRepeat{Do}{do}{while}%
\SetKwComment{Comment}{$\triangleright$\ }{}

\SetCommentSty{mycommfont}

\usepackage{stackengine}
\usepackage{placeins}

\setquotestyle{english}

\usepackage[usenames, dvipsnames]{color}
\addto\captionsenglish{}
\usepackage{subcaption}
 \usepackage{lineno}
\begin{document}
\graphicspath{{./Figures/}}
\begin{frontmatter}



\title{Optimal Adaptive Inspection and Maintenance Planning for Deteriorating Structural Systems}


\author[ERA]{Elizabeth Bismut\corref{cor1}} \ead{elizabeth.bismut@tum.de}
\author[ERA]{Daniel Straub} \ead{straub@tum.de}

\cortext[cor1]{Corresponding author.} % Tel.: +49 89 289 230XX.}

\address[ERA]{Engineering Risk Analysis Group, Technische Universität München.\\ Arcisstraße 21, 80290 München, Germany}

\begin{abstract}

Optimizing inspection and maintenance (I\&M) plans for a large deteriorating structure is a computationally challenging task, in particular if one considers interdependences among its components. This is due to the sheer number of possible decision alternatives over the lifetime of the structure and the uncertainty surrounding the deterioration processes, the structural performance and the outcomes of inspection and maintenance actions. To address this challenge, \cite{Luque_Straub_17} proposed a heuristic approach in which I\&M plans for structural systems are defined through a set of simple decision rules. Here, we formalize the optimization of these decision rules and extend the approach to enable adaptive planning. The initially optimal I\&M plan is successively adapted throughout the service life, based on past inspection and monitoring results. The proposed methodology uses stochastic deterioration models and accounts for the interdependence among structural components. The heuristic-based adaptive planning is illustrated for a structural frame subjected to fatigue.\\
\end{abstract}


\end{frontmatter}
\let\today\relax

\newpage
\section{Introduction}
\label{First}
 Civil and structural assets naturally deteriorate due to mechanisms such as corrosion or fatigue. These can decrease the structural performance and potentially lead to structural failure. Timely interventions on the structure, such as maintenance, repair or replacement of structural components can offset the effects of deterioration and ageing, at a cost to the operator. These intervention costs can represent a significant part of the operation and maintenance budget, especially if not properly planned. As an example, for wind turbines, \citep{Nielsen_Sorensen_10,Roeckmann_et_al_17} report that up to 30\% of the cost of energy is spent on operation and maintenance. These interventions have traditionally been based on industry standards, expert knowledge and empirical studies of the underlying mechanisms of deterioration. An optimal planning is hindered by the large uncertainty associated with deterioration processes. While the scientific literature abounds with stochastic models of deterioration, from corrosion to fatigue \citep[e.g.][]{Southwell_et_al_79,Yang_94,Newman_98,Melchers_03}, in engineering practice deterioration can only be predicted to a limited extent. The main reasons are the variable and uncertain production and environmental factors and material properties \citep{Irving_McCartney_77,Wirsching_Chen_88,King_98,Newman_98, Melchers_03}. I\&M data can reduce this uncertainty and inform future maintenance decisions \citep{Thoft_Sorensen_87,Enright_Frangopol_99, Straub_04,Straub_Faber_05,Straub_Faber_06,Nielsen_Sorensen_15}. Hence, I\&M is an essential part of the structural integrity management.

\begin{figure}[t!]
	\centering
	\includegraphics[trim=130 0 250 0,clip,scale=0.7]{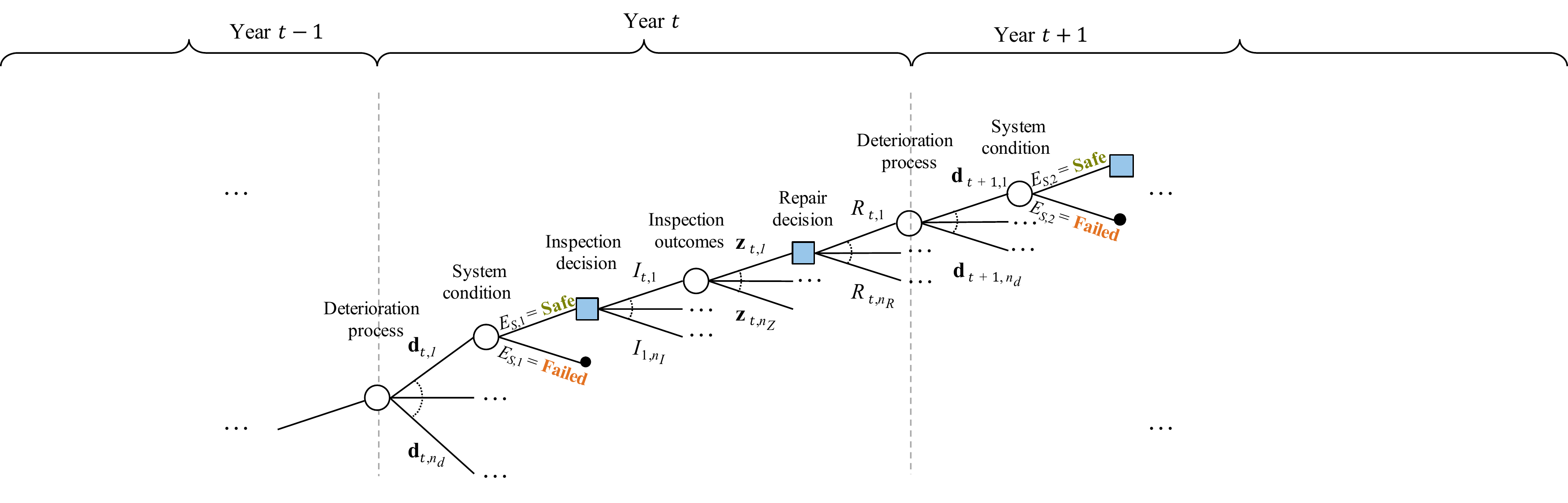}
	\caption{Detail of a decision tree for a deteriorating system which can be inspected and repaired. Random events and outcomes are represented by round nodes, and decisions by square nodes. The deterioration process includes all possible states of damage that the system components can take. Similarly, the inspection outcomes and repair decisions collect all the data and actions for all system components.}\label{Fig:DecisionTree}
\end{figure}

From the perspective of an operator, a good I\&M plan should balance the expected rewards (e.g. increase in system reliability) with the expected I\&M cost over the lifetime of the asset. The identification of the optimal I\&M strategy for a deteriorating structure, also referred to as risk-based inspection (RBI) planning, belongs to the class of stochastic sequential decision problems \citep{Howard_60,Raiffa_Schlaifer_61,Straub_04}. 
The problem is illustrated by the decision tree in Figure~\ref{Fig:DecisionTree}. Each path in this tree corresponds to a sequence of decisions and events, which are associated with a total life-cycle cost and a probability of occurrence. The sequential decision problem consists in finding a set of rules (a policy) at each decision node, which minimizes the expected total life-cycle cost. The decision tree grows exponentially with the considered number of time steps and with the number of system components, and it grows polynomially with the number of available actions and with the number of deterioration states and observations.

Numerous algorithms have been proposed towards the solution of this general optimization problem.
The founding algorithms were developed in the 1950s in the field of operations research (OR) to tackle a large variety of problems, from optimizing the location of warehouses in order to minimize transportation costs, to the management of a baseball team \citep{Howard_60}.
Bellman and Howard described solution strategies for fully observable Markovian decision processes (MDPs), based on Bellman's Dynamic Programming tool \citep{Bellman_57a,Bellman_57b,Howard_60}. \cite{Raiffa_Schlaifer_61} formally included the uncertainty associated with the information acquired during the decision process, paving the way to the study of partially observable MDPs (POMDPs) \citep{Astrom_65,Kaelbling_et_al_98}.

The POMDP approach attempts to provide a universal plan for decision problems where the process is Markovian, and, in theory, accounts for all possible scenarios \citep{Kochenderfer_15}. \citep{Papadimitriou_Tsitsiklis_87} showed that the exact solution of the finite-horizon POMDP cannot be found in polynomial time and is indeed PSPACE-complete. \citep{Bellman_57a} coined the term \emph{curse of dimensionality} and \cite{Pineau_et_al_06} added the term \emph{curse of history} to describe this complexity. Approximate and tractable offline and online solutions for POMDPs using point-based value iteration solvers and belief state approximations are available and have been applied to selected I\&M planning problems \citep{Durango_Madanat_02,Papakonstantinou_14,Nielsen_Sorensen_15,Memarzadeh_Pozzi_16,Schoebi_Chatzi_16,Papakonstantinou_et_al_18}. 
These approaches perform well where the dimensionality of the problem remains moderate. However, they scale poorly for large multi-component systems, which can easily have a state-space of size $10^{100}$ or larger \citep{Bismut_et_al_17}. 

An alternative to classical POMDP solvers is a direct policy search \citep{Ng_et_al_00,Rosenstein_Barto_01,Powell_11}, in which one aims at identifying directly and jointly the optimal set of rules for all decisions. This is analogous to the \textit{normal form} of the decision analysis described by \cite{Raiffa_Schlaifer_61}. \cite{Sutton_et_al_00} have shown that direct policy search implemented through policy-gradient methods present algorithmic and convergence advantages over value-iteration-based POMDP solvers. Direct policy search does not require Markovian assumptions on the model, nor the explicit computation of the belief states. Direct policy search is also the basis for policy search algorithms using deep neural networks. Popular in the field of motion planning and robotics \citep[e.g.][]{ Chebotar_et_al_17}, these algorithms have recently been adapted to I\&M planning and address the high dimensionality of the problem \citep{Andriotis_Papakonstantinou_19}.

In the context of I\&M planning, the engineering understanding allows the identification of suitable functional forms of the policies. We call these \textit{heuristics}. For example, a simple heuristic is to perform an inspection campaign whenever the reliability of the structure falls below a threshold. Through such heuristics, the implementation of a direct policy search is straightforward. The optimization is performed over a few heuristic parameters to search the space of policies (Section \ref{Section:DPS_off} expands on this methodology). In the context of optimizing inspections for structural components, it has been found that heuristics lead to solutions that are very close to the optimum POMDP solutions in terms of the resulting expected total life-cycle costs \citep{Nielsen_Sorensen_15}.

Most RBI planning methodologies are based on such heuristic strategies, mostly in an ad-hoc manner \citep[e.g.][]{Thoft_Sorensen_87,Lam_Yeh_94,Faber_et_al_00,Straub_04,Moan_05,Nielsen_Sorensen_14}. These methodologies perform the optimization component-by-component, without considering the interaction among components. Hence it cannot be ensured that the resulting plans are optimal at the system level. In fact, \citep{Luque_Straub_17} show that component-based optimization can lead to strongly sub-optimal I\&M plans. \citep{Straub_04, Straub_Faber_05} consider an extension to address system effects in a simplified manner. Other studies have proposed a heuristic-based maintenance planning for series and parallel systems with independent or fully correlated failure modes, but do not account for the effect of inspection results on the components and system reliability \citep{Barone_Frangopol_14}.

\citep{Luque_Straub_17} extend the heuristic approach to a system-level I\&M planning, which accounts for the interdependence among components. The methodology relies on the computation of the system reliability by means of a dynamic Bayesian network (DBN) model developed in \citep{Straub_09, Luque_Straub_16}. They evaluate the expected cost of selected system-level I\&M plans without performing a formal optimization. Here, we extend the methodology by optimizing the plans by means of the cross entropy method \citep{Rubinstein_Kroese_17}. The I\&M plans are evaluated in terms of expected total life-cycle I\&M cost and failure risk. Heuristic parameters are proposed that prescribe inspection times and locations.

A major contribution of this paper is the introduction of adaptive planning, whereby the heuristic I\&M plan is modified as new information through inspections and monitoring becomes available. We show that re-optimizing the heuristic parameters once new information is available leads to better I\&M plans. 

Section~\ref{Section:Methodology} summarizes the general methodology for heuristic optimization and introduces the adaptive approach. In Section~\ref{Section:ExpectedCost}, we first address the computation of the expected total life-cycle cost and then propose the use of the cross-entropy method for solving the optimization problem. In Section~\ref{Section:Heuristic}, we discuss the choice of heuristics for I\&M planning in structural systems. Section~\ref{Section:Numerical} presents the application of the methodology to a steel structure subjected to fatigue that includes correlation among components. It is followed by results and discussion in Sections~\ref{Section:Results} and~\ref{Section:ConcludingRemarks}.

\section{Optimal adaptive heuristic planning}
\label{Section:Methodology}
\subsection{The generic strategy optimization problem}
The premises of the problem are summarized in the following.
\begin{enumerate}[label=(\alph*)]
    \item We assume that a model describing the dynamics of the system (deterioration, loads, structural response), including a prior probabilistic model of uncertain parameters, is available to the analyst. The state of the system is represented by random variables $\bm{\Theta}$, which include the time varying system capacity and the applied loads.
	\item We consider a discrete-time model with a fixed finite-horizon $T$, which is typically the anticipated service life of the system. A finite service life is a reasonable assumption since the durability of materials, technological advances, changing user demands and requirements are likely to make the structure obsolete eventually. The service life is in general subject to uncertainty; the effect this could have on the results is not further investigated in this paper.  Time is discretized; 
	$i$ indicates the $i^{th}$ time step between times $t_{i-1}$ and $t_i$, where $t_0=0$ is the beginning of the system's service life. $n_T$ denotes the last time step between times $t_{n_T-1}$ and $t_{n_T}=T$. 
	\item A policy $\pi_i$ is the set of rules adopted at time step $i$ guiding the decision process based on the information available at that time \citep{Jensen_Nielsen_07}. When considering a deteriorating multi-component system, the policy takes as input all or part of the current knowledge on the state of the system, and gives the answer to the questions `Inspect?' \{yes, no\}, `Where?' \{component $j,~ k$, ...\},`What to look for?' \{corrosion, fatigue,...\} ,`How?' \{visually, ultrasonic inspection, thickness measurements, ...\}\textbf{}, `Repair?' \{yes, no, how\}. The system knowledge includes the history of inspection outcomes, monitoring data, repairs and component failures. A policy is \textit{stochastic} if it assigns an action following a probability distribution; it is otherwise \textit{deterministic}. Here we consider deterministic policies, but the general methodology is applicable to either type of policies.
	\item A strategy $\mathcal{S}=\{\pi_1, \pi_2,...\pi_{n_T}\}$ is the set of policies for all time steps. $\mathscr{S}$ is the space of all strategies. A strategy is \textit{stationary} if its policies are identical for all time steps, i.e. $\pi_1=\pi_2=...=\pi_{n_T}=\pi$. 
	\item A heuristic parametrizes a strategy. A heuristic strategy is denoted by $\mathcal{S}_{\bm{w}}$ and is governed by heuristic parameters $\bm{w}=\{w_1,w_2,...w_h\}$.
	\item $\bm{Z}=\{\bm{Z}_{1},...,\bm{Z}_{n_T}\}$ are random vectors describing inspection and monitoring outcomes collected during the service life of the structure. $\bm{Z}_{1:i}$ denotes the vector of outcomes collected up to time step $i$.
	\item $N$ is the number of system components considered for inspection. The components are typically indexed by $k$.
\end{enumerate}
The optimization of the sequential decision process aims to identify the decision strategy $\mathcal{S}^*$ that minimizes the expected total life-cycle cost:
\begin{equation} \label{Eq:Opt}
\mathcal{S}^*=\underset{\mathcal{S}\in{\mathscr{S}}}{\arg \min}~\mathbf{E}[C_{\text{tot}}|\mathcal{S}],
\end{equation}
where $C_{\text{tot}}$, also written as $C_{\text{tot},1:{n_T}}$, is the discounted total life-cycle cost, including inspections, repairs, and possible failures of the system occurring within the time horizon $T$. $\mathbf{E}[C_{\text{tot}}|\mathcal{S}]$ is the expected total life-cycle cost when strategy $\mathcal{S}$ is implemented. Its computation is described in Section~\ref{Section:ExpectedCost}.
The next section summarizes the heuristic approach to solving the optimization problem of Equation~\ref{Eq:Opt} and thereafter introduces the adaptive heuristic planning. 

\subsection{Direct policy search and adaptive search}\label{Section:DPS}
\subsubsection{Direct policy search with heuristic parameters $\bm{w}$} \label{Section:DPS_off}
First consider a system with $N$ components that can be inspected, then potentially repaired, at every time step $i$. There are three possible courses of action: either one does nothing, or inspects and based on the inspection result either repairs or does not repair. 
The total number of possible actions at each time step is $n_{actions}=3^N$. All observations from inspection and monitoring during a time step $i$ are summarized in $\bm{Z}_i$, which has a discrete outcome space of size $n_{obs}$. In the simplest case, the inspection outcome for each component  is either no inspection, or inspection and no detection, or inspection and detection, in which case $n_{obs}=3^N$.

A deterministic policy $\pi_i$ at time step $i$ chooses an action among the $3^N$ options, in function of the observation history, which can take $n_{obs}^i$ distinct realizations. Hence, the total number of potential policies at time step $i$ is $3^{N{n_{obs}^i}}$. It follows that there are $\prod_{i=1}^{n_T} 3^{N{n_{obs}^i}}={3}^{N\frac{n_{obs}^{n_T}-1}{n_{obs}-1}}$ distinct strategies. This illustrates how the space of strategies $\mathscr{S}$ increases exponentially with the number of time steps $n_T$ and the number $N$ of system components. 

Going through each of the $3^{Nn_T}n_{obs}^\frac{n_T^2+n_T}{2}$ strategies is clearly impossible. Instead, the solution space $\mathscr{S}$ is reduced by choosing a suitable \textit{heuristic}, parametrized by $\bm{w}=\{w_1,w_2,...w_h\}$. A simple (although likely suboptimal) example of a heuristic is: Inspect all components of the structure whenever the reliability estimate is below a threshold, then repair all components at which defects are found. In this case, the only heuristic parameter is the reliability threshold. The strategies $\mathcal{S}_{\bm{w}}$  resulting from such a heuristic form a subspace of $\mathscr{S}$.

The solution $\mathcal{S}^*$ to the optimization problem in Equation~\ref{Eq:Opt} is therefore approximated by $\mathcal{S}_{\bm{w}_0^*}$, where
\begin{equation}\label{Eq:Heuristic}
\bm{w}_0^*=\underset{\bm{w}}{\arg\min}~\mathbf{E}\left[C_{\text{tot}}|\mathcal{S}_{\bm{w}}\right].
\end{equation}
From this point forward, we use $\mathbf{E}[\cdot|\bm{w}]$ to denote $\mathbf{E}[\cdot|\mathcal{S}_{\bm{w}}]$ for the sake of simplicity in the notation.

Restricting the optimization of Equation~\ref{Eq:Opt} to the heuristic space of strategies has two effects: firstly, it reduces the solution space; secondly, the heuristic parameters, such as a reliability threshold, can take continuous values.

One challenge is the selection of the heuristic, so that the strategies explored are close enough to the exact solution of Equation~\ref{Eq:Opt}. In the context of machine learning, this selection is controlled by hyperparameters that can in turn be optimized, or "tuned". 
This tuning aspect is not investigated further in the paper. 

For I\&M problems, the choice of heuristics is often driven by operational constraints, such as the need for regular inspection intervals. The heuristic can also include reliability criteria that need to be fulfilled \citep{Tsang_95, Nielsen_Sorensen_14}. The heuristics are also not required to incorporate information about the system. For instance, a valid strategy is one that prescribes to replace a system every 10 years, regardless of its actual state of deterioration or of any prior collected information. Such systematic plans are typically adopted for non-crucial components of a system (e.g. air filters of an air handling unit); they are rarely based on a quantitative optimization, however. We introduce heuristics for I\&M planning in deteriorating structures in Section \ref{Section:Heuristic}.

\subsubsection{Adaptive policy search}\label{Section:Adaptive}
Following Equation~\ref{Eq:Heuristic}, the optimal heuristic is found initially and kept throughout the system lifetime $T$. However, as new observations $\bm{Z}_{1:i}$ become available, the initially optimal strategy may no longer be optimal. Therefore, we suggest to adapt the strategy during the lifetime of the structure by adding an on-line computation that accounts for the new observations $\bm{Z}_{1:i}$.

Initially, one performs the optimization with the prior model following Equation~\ref{Eq:Heuristic} to obtain the optimal parameters $\bm{w}_0^*$. To make this explicit, the actions (e.g. inspections and eventual repairs) are performed as dictated by $\bm{w}_0^*$ until a time step $j_1$, typically when new information is available. At $t_{j_1}$, the decision maker has the opportunity to improve the strategy. 

By updating the prior model with information obtained up to time $t_{j_1}$, $\bm{Z}_{1:j_1}$, a heuristic strategy optimization is again performed for the rest of the service life of the structure, and a new parameter value $\bm{w}_1^*$ and its associated strategy are obtained (Equation~\ref{Eq:2}).
\begin{equation}\label{Eq:2}
\bm{w}_{1}^*=\underset{\bm{w}}{\arg\min}~\mathbf{E}\left[C_{\text{tot},j_1:n_T}  |\bm{w},\bm{Z}_{1:j_1} \right].	
\end{equation}
$C_{\text{tot},j_1:n_T}$ is the total cost evaluated from time step $j_1$ onwards, as opposed to $C_{\text{tot},1:n_T}$, evaluated from time step $1$ onwards. $\mathbf{E}\left[\cdot|{\bm{w},\bm{Z}_{1:j_1}}\right]$ is the expectation operator conditional on the observation and repair history up to time step $j_1$, applying strategy parameters $\bm{w}$.

It follows from Equation~\ref{Eq:2} that
\begin{equation}\label{Eq:6}
\mathbf{E}\left[C_{\text{tot},j_1:n_T}  |\bm{w}_0^*,\bm{Z}_{1:j_1}\right]\geq \mathbf{E}\left[C_{\text{tot},j_1:n_T}  |\bm{w}_{1}^*,\bm{Z}_{1:j_1} \right].
\end{equation}

\begin{figure}[t!]
	\centering
	\includegraphics[scale=0.6]{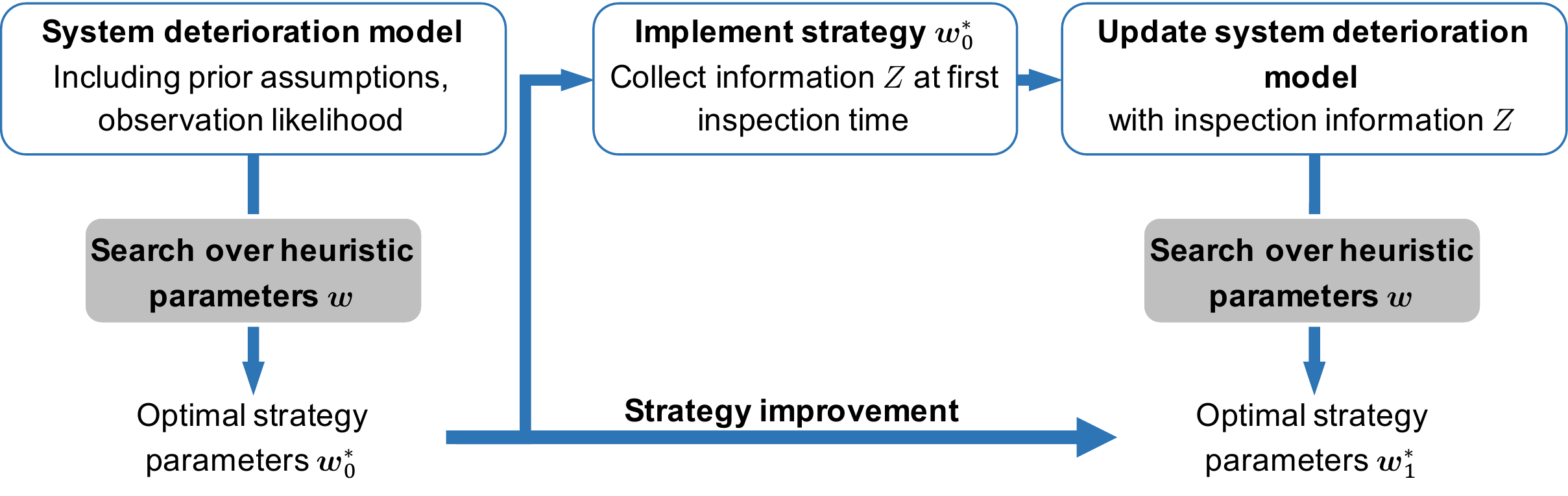}
	\caption{One-step adaptive I\&M planning with heuristic search}\label{Fig:Flowchart_Adaptive}
\end{figure}

Our aim is to show that the adaptive planning decreases the expected total life-cycle cost. For this purpose, we decompose the expected total life-cycle cost of strategy $\bm{w}_0^*$ into
\begin{equation}\label{Eq:3}
\mathbf{E}\left[C_{\text{tot},1:n_T} |\bm{w}_0^*\right]= \mathbf{E}\left[C_{\text{tot},1:j_1} |\bm{w}_0^*\right]+\mathbf{E}_{\bm{Z}_{1:j_1}} \left[\mathbf{E} \left[C_{\text{tot},j_1:n_T}  |\bm{w}_0^*,\bm{Z}_{1:j_1}\right]\right],
\end{equation}
where $C_{\text{tot},1:\tau_1} |\bm{w}_0^*$ is the cost incurred until time $t_{j_1}$ following strategy $\mathcal{S}_{\bm{w}_0^*}$. The expectation $\mathbf{E}_{\bm{Z}_{1:j_1}}[\cdot]$ operates on the observation and repair history up to time step $j_1$. 
 This operator preserves the inequality in Equation~\ref{Eq:6}, hence by combining Equations \ref{Eq:6} and \ref{Eq:3}, we obtain that
\begin{equation}\label{Eq:4}
\mathbf{E}\left[C_{\text{tot},1:n_T}  |\bm{w}_0^* \right]\geq \mathbf{E} \left[C_{\text{tot},1:j_1} |\bm{w}_0^*\right]
+\mathbf{E}_{\bm{Z}_{1:j_1} } \left[\mathbf{E}\left[C_{\text{tot},j_1:n_T } |\bm{w}_{1}^*,\bm{Z}_{1:j_1} \right]\right].
\end{equation}
The difference between the left and right hand side of Equation \ref{Eq:4} quantifies the expected gain by adapting the strategy at time $t_{j_1}$. Figure~\ref{Fig:Flowchart_Adaptive} retraces the steps of this strategy improvement.

In the context of production and supply optimization, a similar approach is known as model predictive control (MPC)  \citep{Pooya_Pakdaman_18}. The principle is also adopted to improve reinforcement learning algorithms \citep{Kahn_et_al_17}, and is key to the Monte Carlo Tree Search POMCP algorithm presented in \citep{Silver_Veness_10}.

The strategy adaptation can be repeated every time new information is collected. The array $\bm{W}^*$ stores the successively adapted heuristic parameters values:
\begin{equation}\label{Eq:5}
\bm{W}^*=\{\bm{w}_0^*,\bm{w}_1^*,\bm{w}_2^*,…,\bm{w}_{n_{ADAPT}}^*\},
\end{equation}
Algorithm \ref{Alg:Adapt} summarizes the adaptive planning method.

\begin{algorithm}[h!]
	\SetAlgoLined
	\SetKwInOut{Input}{input}\SetKwInOut{Output}{output}
	\Input{heuristic, number of times steps $n_T$ for time horizon $T$, \\adaptation times $\bm{j}=\{j_1,\dots,j_{n_{ADAPT}}\}$, $prior\_model$}
	\Output{$\bm{W}^*$}
	$l\leftarrow0$\;
	$model \leftarrow prior\_model$\;
	
	\Repeat{$l> n_{ADAPT}$}{
		$\bm{w}^*_l\leftarrow \underset{\bm{w}}{\arg\min}(\mathbf{E}[C_{\text{tot},j_{l+1}:n_T }|\bm{w},model])$ \Comment*[r]{\def\stackalignment{l}{\stackanchor{find optimal heuristic parameters,} {e.g. with Algorithm~\ref{Alg:2}}}}
		$l\leftarrow l+1$\;
		\If{$\text{\upshape no system failure before}~t_{j_l}$}{
		   Perform inspections and repairs following $\bm{w}_{l-1}^*$ until $t_{j_l}$ 
		    
		    $model\leftarrow update\_model(model, \bm{z_{1:j_l}})$\Comment*[r]{update deterioration model with $\bm{z}_{1:j_l}$}
		}
		
	}
	\Return{$\bm{W}^*=\{\bm{w}_0^*,...,\bm{w}_{n_{ADAPT}}^*\}$}
	\caption{Procedure for adaptive I\&M planning}
	\label{Alg:Adapt}
\end{algorithm}

The proposed adaptive planning is an on-line optimization at the level of array $\bm{W}^*$ and is greedy since the strategy optimization is always performed assuming that it is the last opportunity to optimize the strategy. Multi-step-based improvement greedy techniques have demonstrated to perform much better than a single initial optimization, as shown empirically with a number of  algorithms \citep{Silver_Veness_10,Pooya_Pakdaman_18, Efroni_et_al_18}.

\section{Stochastic computation and optimization}\label{Section:ExpectedCost}
\subsection {Expected total cost of a strategy}\label{Section:CostCalc}
\subsubsection{Cost breakdown}
\begin{figure}[t!]
	\centering
	\includegraphics[scale=0.5]{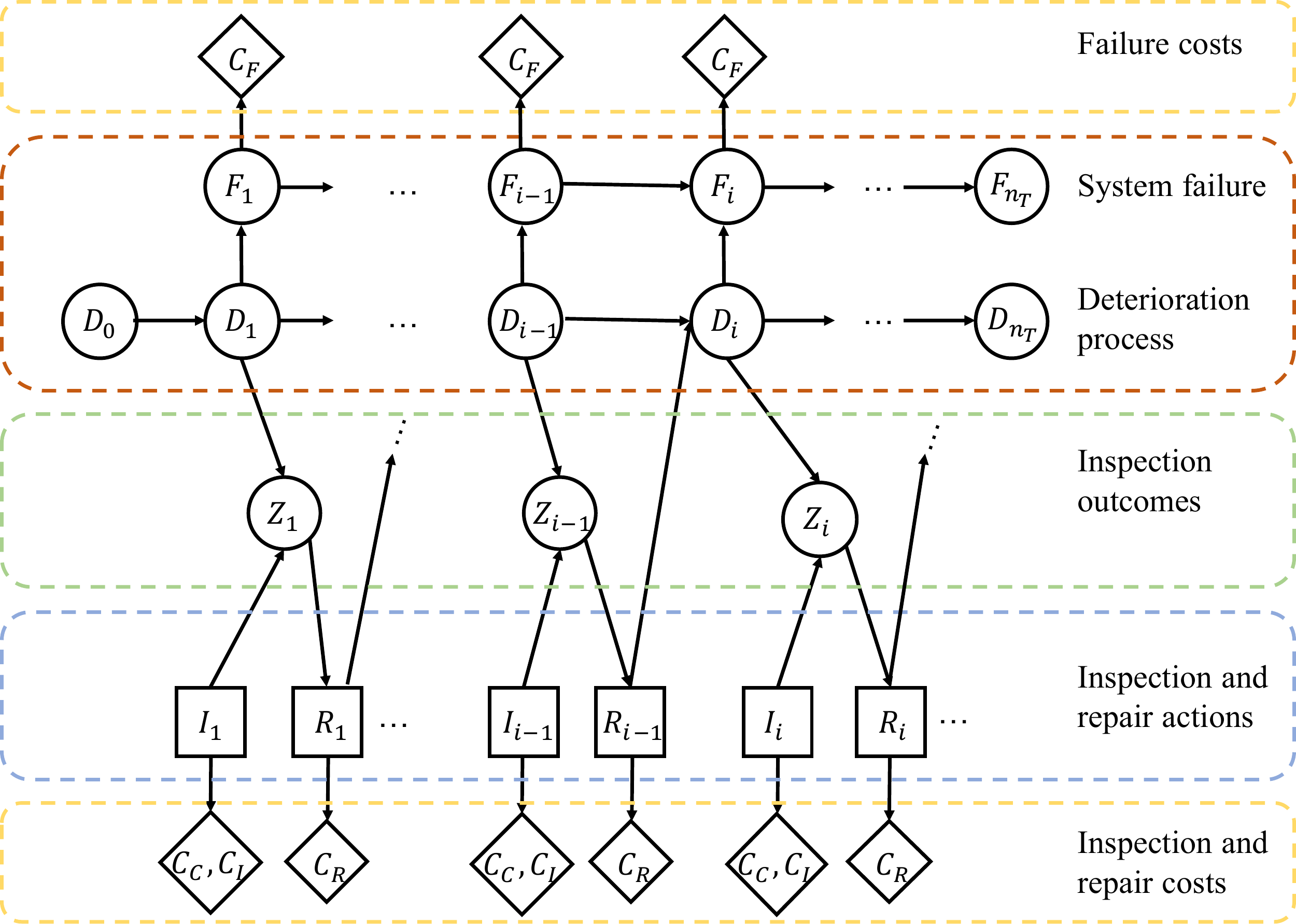}
	\caption{Influence diagram for the decision process. The square nodes indicate the inspection and repair decisions, and the diamond shaped nodes the inspection, repair and failure costs.}\label{Fig:ID}
\end{figure}

The heuristic approach requires the computation of the objective function $\mathbf{E}[C_{\text{tot}}|\bm{w}]$ in Equation \ref{Eq:Heuristic}. The influence diagram of Figure \ref{Fig:ID} shows all costs incurred during the service life of the structure.

The expectation $\mathbf{E}[C_{\text{tot}}|\bm{w}]$ of the total cost of a strategy operates on the state of the system $\bm{\Theta}$, which includes the time varying system capacity and the applied loads, and on the observation outcomes $\bm{Z}\in \Omega_{\bm{Z}}$.
\begin{equation} \label{Eq:Expected}
\mathbf{E}[C_{\text{tot}}|\bm{w}]
=\mathbf{E}_{\bm{\Theta},\bm{Z}} [C_{\text{tot}} |\bm{w}]
=\int_{\Omega_{\bm{\Theta}}}\int_{\Omega_{\bm{Z}}}
{C_{\text{tot}}(\bm{w},\mathbf{z},\bm{\theta}) f_{\bm{\Theta},\bm{Z}|\bm{w}}(\bm{\theta},\mathbf{z}|\bm{w})\mathrm{d}\mathbf{z}\mathrm{d}\bm{\theta}},
\end{equation}
where $f_{\bm{\Theta},\bm{Z}|\bm{w}}(\bm{\theta},\mathbf{z}|\bm{w})$ is the joint probability distribution of $\bm{\Theta}$ and $\bm{Z}$ conditional on the strategy parameters $\bm{w}$.

This expectation can be approximated with crude Monte Carlo simulation (MCS). However, when the expected costs involve the risk of failure, and if the underlying probability of failure is small, then the Monte Carlo sampling over both $\bm{\Theta}$ and $\bm{Z}$ is computationally expensive because a large number of samples is necessary to achieve an acceptable accuracy. Therefore, following \citep{Luque_Straub_17} we propose to first evaluate the expected costs conditional on $\bm{Z}$ (Equation~\ref{Eq:TotalCost}), and then perform a MCS over the observation history $\bm{Z}$. To this end, Equation~\ref{Eq:Expected} is rewritten as
\begin{equation} \label{Eq:Expected_2}
\mathbf{E}[C_{\text{tot}}|\bm{w}]
=\int_{\Omega_{\bm{Z}}}{\mathbf{E}_{\bm{\Theta}|\mathbf{z}} [C_{\text{tot}}(\bm{w},\mathbf{z},\bm{\Theta})|\bm{w},\mathbf{z}]f_{\bm{Z}|\bm{w}}(\mathbf{z}|\bm{w})\mathrm{d}\mathbf{z}},
\end{equation}
where $f_{\bm{Z}|\bm{w}}(\mathbf{z}|\bm{w})$ is the probability density function of the service life observations $\bm{Z}$ conditional on $\bm{w}$.
The corresponding Monte Carlo estimate is detailed in Section~\ref{Subsection:MCS}.

The total life-cycle cost $C_{\text{tot}}(\bm{w},\bm{Z},\bm{\Theta})$ is the sum of the inspection campaign costs $C_C(\bm{w},\bm{Z},\bm{\Theta})$, component inspection costs $C_I(\bm{w},\bm{Z},\bm{\Theta})$, repair costs $C_R(\bm{w},\bm{Z},\bm{\Theta})$ and failure costs $C_F(\bm{w},\bm{Z},\bm{\Theta})$ over the lifetime of the structure. All these costs are present values, i.e., they are discounted to time $0$.
Hence the expected total life-cycle cost conditional on the observation outcomes $\bm{Z}$ for given heuristic parameters $\bm{w}$ is 
\begin{equation} \label{Eq:TotalCost}
\mathbf{E}_{\bm{\Theta}|\bm{Z}} [C_{\text{tot}}|\bm{w},\bm{Z}]=\mathbf{E}_{\bm{\Theta}|\bm{Z}} [C_I|\bm{w},\bm{Z}]+\mathbf{E}_{\bm{\Theta}|\bm{Z}} [C_C|\bm{w},\bm{Z}]+\mathbf{E}_{\bm{\Theta}|\bm{Z}} [C_R|\bm{w},\bm{Z}]+\mathbf{E}_{\bm{\Theta}|\bm{Z}} [C_F|\bm{w},\bm{Z}].
\end{equation}
The following sections detail how the individual terms in Equation \ref{Eq:TotalCost} are evaluated.

\subsubsection{Computing the conditional risk of failure}\label{Subsection:RoF}
Failure is here considered a terminal event. This means that if failure of the system occurs before the end of the service life, no further inspection and repair actions are explicitly considered from that point on. This simplifying assumption does not significantly affect the estimate of the expected total life-cycle cost, due to the high reliability of infrastructure systems \citep{Kuebler_Faber_04}. The fixed cost  $c_F$ in case of failure includes replacement costs and future life-cycle costs of the new structure. When failure occurs at time $t$, the associated cost expressed as present value is $\gamma(t)\cdot c_F$, where $\gamma(t)$ is the discount factor. 

The conditional risk of failure over the lifetime of the structure $\mathbf{E}_{\bm{\Theta}|\bm{Z}} [C_F|\bm{w},\bm{Z}]$ can be defined in terms of $T_F$, the time to failure of the system. $T_F$ is a random variable, and is dependent on the implemented strategy $\mathcal{S}_{\bm{w}}$ and the system history $\bm{Z}$. Its probability distribution function (pdf) is denoted by $f_{T_F |\bm{w},\bm{Z}} (t)$. It is
\begin{equation} \label{Exp_TTF}
\mathbf{E}_{\bm{\Theta}|\bm{Z}} [C_F|\bm{w},\bm{Z}]=\mathbf{E}_{T_F |\bm{Z}} [C_F|\bm{w},\bm{Z}]=\int_{0}^{T}{c_F\cdot \gamma(t)\cdot f_{T_F |\bm{w},\bm{Z}} (t)\mathrm{d}t}.
\end{equation}
As time is discretized in $n_T$ time steps (years), the integration in Equation \ref{Exp_TTF} is approximated by
\begin{equation}  \label{Eq:Exp_Risk}
\mathbf{E}_{\bm{\Theta}|\bm{Z}} [C_F|\bm{w},\bm{Z}]\simeq c_F\cdot\sum_{i=1}^{n_T}{\gamma(t_i )\cdot \left[F_{T_F|\bm{w},\bm{Z}}(t_i)-F_{T_F|\bm{w},\bm{Z}}(t_{i-1})\right]},
\end{equation}
where 
%$F_i$ is the event 'failure of the system before time $t_i$', 
$F_{T_F|\bm{w},\bm{Z}}$ is the cumulative distribution function (CDF) of the conditional $T_F$. In particular, $[F_{T_F|\bm{w},\bm{Z}}(t_i)-F_{T_F|\bm{w},\bm{Z}}(t_{i-1})]$ is the annual probability of failure for year $i$. We compute this annual probability of failure conditional on the observation outcomes with the DBN model described in Section~\ref{Subsection:DBN}.

\subsubsection{Computing the conditional expected inspection and repair costs}
Since failure is a terminal event, the evaluations of expectations $\mathbf{E}_{\bm{\Theta}|\bm{Z}}[C_I|\bm{w},\bm{Z}]$, $\mathbf{E}_{\bm{\Theta}|\bm{Z}}[C_C|\bm{w},\bm{Z}]$ and $\mathbf{E}_{\bm{\Theta}|\bm{Z}}[C_R|\bm{w},\bm{Z}]$ consider that an observation or repair action at a time $t_i$ can occur only if the system has survived until that time. For instance, the conditional expected life-cycle inspection campaign costs are calculated as
\begin{equation} \label{Eq:CampCosts}
\mathbf{E}_{\bm{\Theta}|\bm{Z}}[C_I|\bm{w},\bm{Z}]=\sum_{i=1}^{n_T}{\gamma(t_i )\cdot c_I(t_i, \bm{w},\bm{Z})\cdot \left[ 1-F_{T_F|\bm{w},\bm{Z}}(t_i) \right]},
\end{equation}
where $c_I(t_i, \bm{w},\bm{Z})$ is the inspection campaign cost incurred at time $t_i$ as prescribed by strategy $\bm{w}$ and inspection history $\bm{Z}$, and $1-F_{T_F|\bm{w},\bm{Z}}(t_i)$ is the probability of survival of the system up to time $t_i$.

Similarly, the conditional expected life-cycle component inspection cost and life-cycle repair cost are evaluated as
\begin{equation} \label{Eq:CompCosts}
\mathbf{E}_{\bm{\Theta}|\bm{Z}}[C_C|\bm{w},\bm{Z}]=\sum_{i=1}^{n_T}{\gamma(t_i )\cdot c_C(t_i, \bm{w},\bm{Z})\cdot \left[1-F_{T_F|\bm{w},\bm{Z}}(t_i)\right]},
\end{equation}
\begin{equation} \label{Eq:RepCosts}
\mathbf{E}_{\bm{\Theta}|\bm{Z}}[C_R|\bm{w},\bm{Z}]=\sum_{i=1}^{n_T}{\gamma(t_i )\cdot c_R(t_i, \bm{w},\bm{Z})\cdot \left[1-F_{T_F|\bm{w},\bm{Z}}(t_i)\right]},
\end{equation}

\subsubsection{Monte Carlo simulation over inspection history $\bm{Z}$}\label{Subsection:MCS}
In Equation \ref{Eq:Expected}, the conditional expectation $\mathbf{E}_{\bm{\Theta}|\bm{Z}} [C_{\text{tot}}|\bm{w},\bm{Z}]$ must be integrated over all possible outcomes $\bm{Z}$.
The integral over $\bm{Z}$ cannot be easily computed analytically, as the probability density of all possible inspection outcomes $f_{\bm{Z}|\bm{w}} (\mathbf{z})$ is not readily available. 

 However, sampling from this distribution is possible, by first generating deterioration histories from the model, and then generating inspection outcomes $\mathbf{z}^{(q)}$ conditional on the deterioration and the adopted strategy. In the case of the adaptive optimization presented in Section \ref{Section:Adaptive}, the deterioration histories have to be generated conditional on the past inspection outcomes. It is then necessary to resort to a Bayesian sampling method. Here we employ the BUS methodology with SuS \citep{Straub_Papaioannou_15}.   
 Equation \ref{Eq:Expected} can then be approximated by MCS:

\begin{equation}\label{Eq:MCS}
\mathbf{E}[C_{\text{tot}}|\bm{w}]\simeq\frac{1}{n_{MC}}  \sum_{q=1}^{n_{MC}}\mathbf{E}_{\bm{\Theta}|\mathbf{z}^{(q)}} [C_{\text{tot}}|\bm{w},\mathbf{z}^{(q)}].
\end{equation}

Notably, this MCS over $\bm{Z}$, conditional on a strategy $\mathcal{S}_{\bm{w}}$, does not need to generate samples that lead to the failure of the system. The probability of failure is included in the risk of failure $\mathbf{E}_{\bm{\Theta}|\mathbf{z}^{(q)}} [C_F|\bm{w},\mathbf{z}^{(q)}]$, which is calculated directly conditional on each sample history $\mathbf{z}^{(q)}$ as per Equation~\ref{Eq:Exp_Risk}. \ref{App:Variance} discusses the accuracy of MCS to evaluate the risk of failure. In numerical investigations, it was found that a quite small number of samples is sufficient to obtain a good approximation of the expected  total life-cycle cost, in the order of 200 samples.

\begin{figure}[t!]
		\centering
			\captionsetup[subfigure]{justification=centering}
\begin{subfigure}[t]{0.5\columnwidth}
	\includegraphics[scale=0.45]{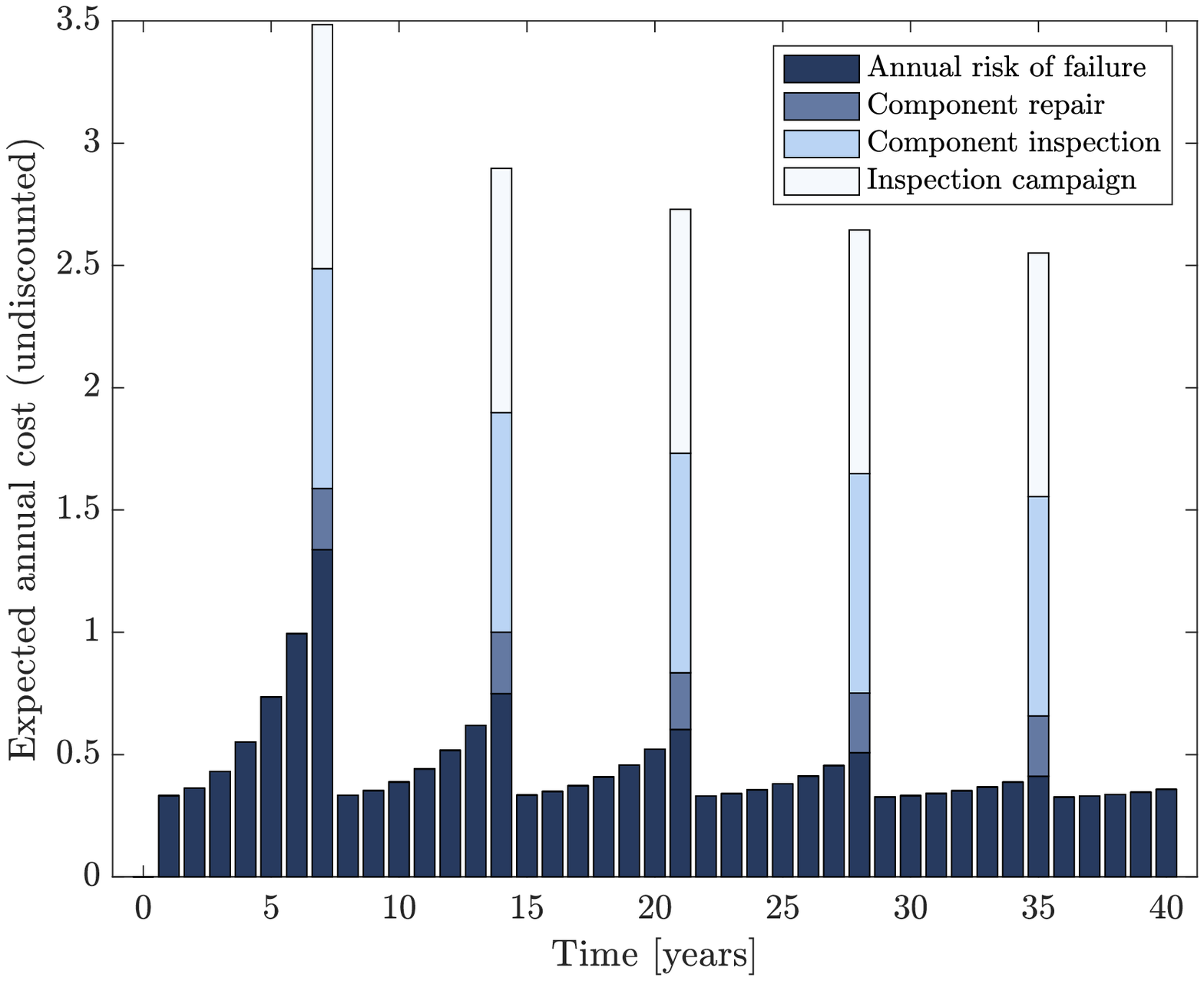}
	\caption{undiscounted}\label{Fig:AnnualCosts_undisc}
\end{subfigure}%
\begin{subfigure}[t]{0.5\columnwidth}
	\centering
	\includegraphics[scale=0.45]{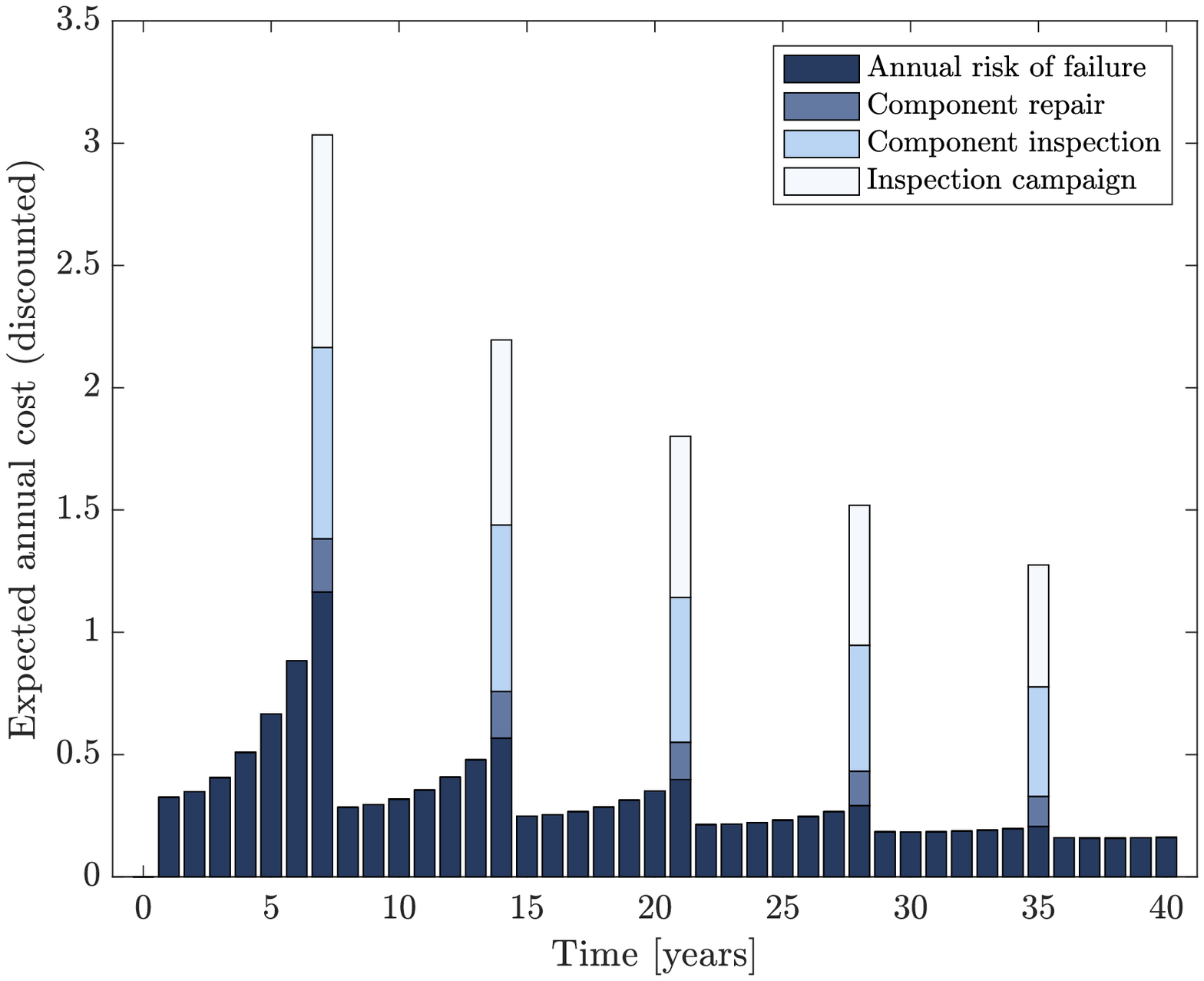}
	\caption{discounted}\label{Fig:AnnualCosts_discount}
\end{subfigure}
\caption{Expected annual costs for strategy $\mathcal{S}_{\bm{w}_{0}^*}$, evaluated with 200 samples, undiscounted (a), and discounted (b) with annually compounded discount rate $r=0.02$.}\label{Fig:AnnualCosts}
\end{figure}

One can break down the expected costs into annual values. Figure~\ref{Fig:AnnualCosts_undisc} shows this breakdown for a selected strategy from the numerical application presented in Section~\ref{Section:Numerical}, evaluated with $n_{MC}=200$ samples. Figure~\ref{Fig:AnnualCosts_discount} shows the effect of the discount factor $\gamma(t)$ on the distribution of the costs for the same strategy. 

\subsection{Conditional reliability of a deteriorating multi-component system}\label{Subsection:DBN}
The computation of the expected cost of failure $\mathbf{E}_{\bm{\Theta}|\bm{Z}} [C_F|\bm{w},\bm{Z}]$ in  Equation~\ref{Eq:Exp_Risk} requires the evaluation of the CDF $F_{T_F|\bm{w},\bm{Z}}$ of the time to system failure, $T_F$.

In this paper, failure of the system originates from the deterioration processes, which gradually decrease the system capacity. Deterioration has been modelled by stochastic processes  \citep[e.g][]{Lin_Yang_85,Noortwijk_09, Shafiee_et_al_15}, or by physics-based equations with uncertain model parameters \citep[e.g.][]{Madsen_et_al_87}. 

The evaluation of expected costs of I\&M strategies requires many reliability evaluations conditional on inspection results. This calls for methods that can efficiently compute conditional reliabilities and can handle a large number of random variables. 

Such efficient computation can be available when using Markov chain models for representing deterioration. In general, deterioration processes are non-Markovian, in that the damaged state is not independent of the past states. However, non-Markovian deterioration processes can be transformed into Markovian processes by state-space augmentation \citep{Straub_09}. The Markov chain model has been extensively used in I\&M research \citep[e.g.][]{Rosenfield_76,Rausand_Amjot_04,Bocchini_et_al_13,Faddoul_et_al_13,Zhu_et_al_13}.

\begin{figure}[t!]
	\centering
	\includegraphics[scale=0.51]{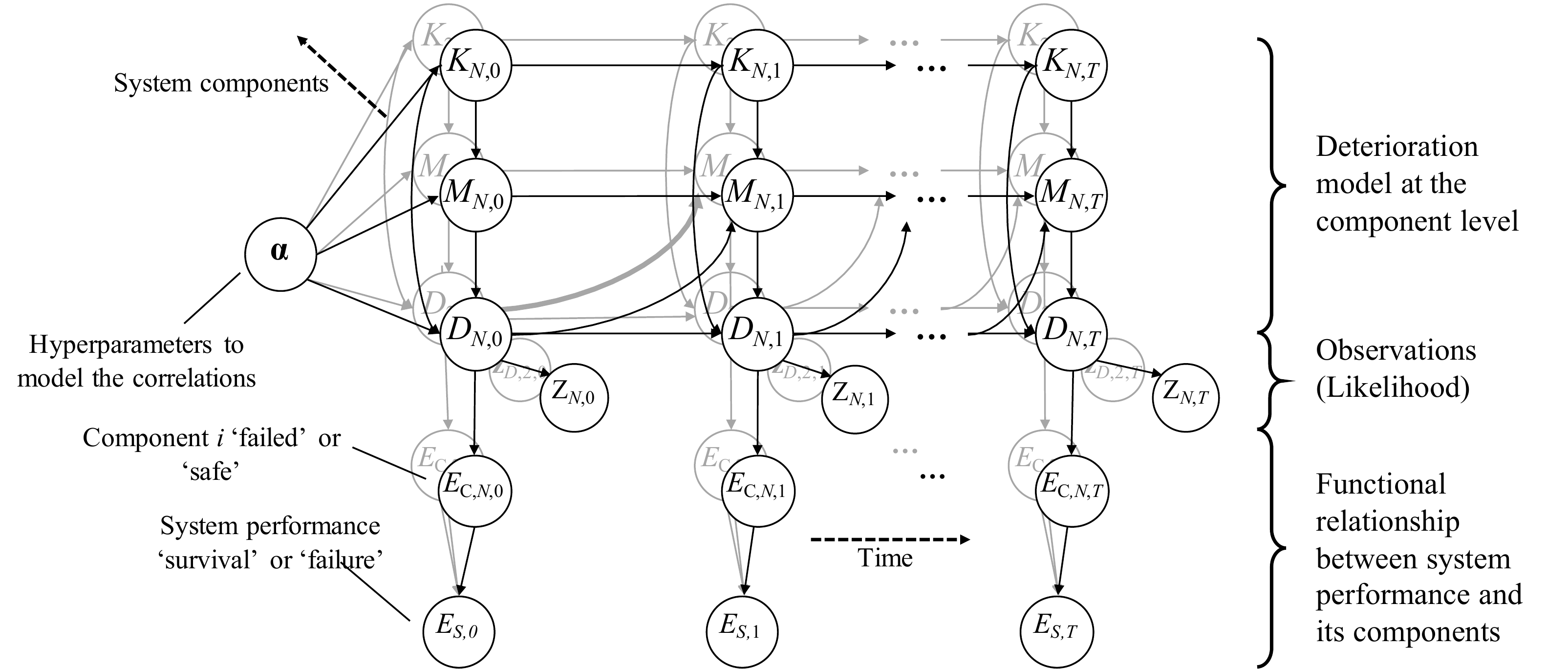}
	\caption{DBN for a deteriorating multi-component system. The correlations between components are explicitly modeled through hyperparameters $\bm{\alpha}$. The state of the system is described by nodes $E_{S,i}$, conditionally on the state of each component, $E_{C,N,i}$, through a function relating components to system state, e.g. a pushover analysis. Evidence is set on the observation nodes $Z_{k,i}$ when the chosen I\&M strategy prescribes it. }\label{Fig:DBN}
\end{figure}

In the majority of real multi-component structural systems, components' deterioration processes are correlated due to common manufacturing conditions and similar environmental and load conditions. To represent deteriorating multi-component systems with interdependent components, standard Markov chain models are not suitable. Instead, we employ the hierarchical dynamic Bayesian network (DBN) of Figure~\ref{Fig:DBN}. This is a variant of the DBN model of \citep{Luque_Straub_17}.

In the model of Figure \ref{Fig:DBN}, each component is represented by its own DBN. These are connected by the hyperparameters $\bm{\alpha}$, which represent the dependence among component deterioration, and the common system performance nodes, which here take states `survival' or `failure'. Observations of deterioration state $D_{k,i}$ of component $0\leq k\leq N$ from inspections or monitoring are represented by $Z_{k,i}$. 
The states of all random variables are discretized according to \citep{Straub_09}. Exact inference algorithms are implemented to evaluate the probability distributions of all random variables conditional on the observations. For details on the modeling and computation in this hierarchical DBN model we refer to \cite{Luque_Straub_16}. 

To improve the computational performance, the nodes $E_{S,i}$ represent the state of the system at time step $i$, if the system does not have to opportunity to fail earlier. This is highlighted by the fact that no arrows link temporally the nodes $E_{S,i}$. We call the event $\{E_{S,i}=\text{`failure'}\}=F_i^*$, the \textit{interval failure event}  \citep{Straub_et_al_18}. 

The evaluation of this DBN results in the conditional probability of $F_i^*$ given the observation from all components.
By introducing the event $F_i$ as `failure of the system up to time $t_i$', we note that $F_i=F_1^*\cup F_2^*\cup…\cup F_i^*$. One can express the CDF $F_{T_F|\bm{w},\bm{Z}}(t_i)$, as
\begin{equation}\label{Eq:CumulativePoF}
F_{T_F|\bm{w},\bm{Z}}(t_i)=\text{Pr}(F_i|\bm{w},\bm{Z})=\text{Pr}(F_1^*\cup F_2^*\cup…\cup F_i^*|\bm{w},\bm{Z}).
\end{equation}

In some systems, the events $F_i^*$ can be assumed to be independent for different $i$. 
In this case, $\text{Pr}(F_i|\bm{w},\bm{Z})$ can be evaluated as
\begin{equation}\label{Eq:ApproxPoF_cumul}
\text{Pr}(F_i|\bm{w},\bm{Z})= 1-\prod_{1\leq j\leq i}{1-\text{Pr}(F_j^*|\bm{w},\bm{Z})}
\end{equation}

The annual probability of system failure for time step $i$ conditional on the observations $\bm{Z}$ is
\begin{equation}\label{Eq:ApproxPoF_annual}
\text{Pr}(F_i|\bm{w},\bm{Z})-\text{Pr}(F_{i-1}|\bm{w},\bm{Z}).
\end{equation}

The DBN model is flexible with respect to the choice of a-priori probability distributions and transition probabilities. The DBN framework, however, has limitations with respect to the number of parameters in the deterioration models, but work-around strategies exist as discussed in \cite{Luque_Straub_16}. 

 The interval probabilities of failure $(\text{Pr}(F_i^*| \bm{w},\bm{Z}))_{0\leq i \leq T}$ are computed with Bayesian inference in the DBN, involving Bayesian filtering, prediction and smoothing algorithms \citep{Askar_Derin_81,Askar_Derin_83,Straub_09,Saarka_13, Luque_Straub_16}. The computational cost of this algorithm increases only linearly with the number of components, which makes it suitable for large systems \citep{Luque_Straub_16}. % (see \ref{App:Complexity}).
 The cumulative probability of failure $\text{Pr}(F_i| \bm{w},\bm{Z})$  and the annual probability of failure are then evaluated with Equations~\ref{Eq:ApproxPoF_cumul} and~\ref{Eq:ApproxPoF_annual}. This implies an approximation, which has shown to be appropriate for the reliability problem investigated in Section~\ref{Section:Numerical} \citep{Straub_et_al_18}.

It should be noted that alternative methods exist for computing $\text{Pr}(F_i|\bm{w},\bm{Z})$. For example, \cite{Schneider_et_al_17} applied BUS with SUS for the same structural system as considered in Section~\ref{Section:Numerical}.

\subsection{Stochastic optimization method}

 Equation~\ref{Eq:MCS} provides a noisy approximation of the expected total life-cycle cost for given heuristic parameters defining a strategy. Stochastic optimization methods are well suited to handle noisy objective functions of the form of Equation~\ref{Eq:MCS} \citep{Spall_12,Hill_13}. 
 Here, we consider the cross entropy (CE) method \citep{DeBoer_et_al_05} combined with a Gaussian process regression (GPR) \citep{Rasmussen_04} to approximate the solution of Equation~\ref{Eq:Heuristic} and to find the heuristic parameters values that minimize the exact expected total life-cycle cost.
 
Algorithm~\ref{Alg:2} summarizes the steps of the CE method, inspired by \citep{Kochenderfer_15}. It generates $n_{CE}$ samples $\bm{w}^{(m)}$ of heuristic parameters from a distribution with parameters $\bm{\lambda}^*$. For each sample $\bm{w}^{(m)}$, $n_{MC}$ observation and monitoring histories are generated and the expected total life-cycle cost is calculated through Equation~\ref{Eq:MCS}. The samples $\bm{w}^{(m)}$ are then ranked in ascending order of their estimated expected life-cycle cost. The $n_{E}$ best-ranked $\bm{w}^{(m)}$, also called elite samples, are used to update the CE sampling distribution parameter $\bm{\lambda}^*$ by cross entropy minimization. This step is repeated until a convergence criterion has been met, or until a sufficient number $n_{max}$ of strategies have been explored. The convergence speed of the CE method can be optimized by choosing $n_{MC}$ adequately. In particular, the CE method can work with a single sample evaluation in Equation \ref{Eq:MCS}, i.e. $n_{MC}=1$, and still converge towards the heuristic parameter values that minimize the exact expected total life-cycle cost \citep{Rubinstein_Kroese_04}. In addition to the CE method, GPR is applied to all the estimated expected costs calculated during the CE loop. The surrogate function thus obtained is used to approximate $\mathbf{E}[C_{\text{tot}}|\mathcal{S}_{\bm{w}}]$ in Equation~\ref{Eq:Heuristic}, and the optimal heuristic parameter values $\bm{w}^*$ are obtained by minimizing this surrogate function. 

Coupling the CE method with GPR presents the advantage that the surrogate mean and standard error of the mean can be extracted for any point in the heuristic parameter space. This can be used to evaluate the sensitivity of the expected total life-cycle cost to the heuristic parameters. Furthermore, due to the CE sampling method, the standard error of the mean decreases towards the surrogate minimum. This methodology can handle both continuous or discrete heuristic parameters, as illustrated in Section~\ref{Section:Numerical}.

\begin{algorithm}[t!]
	\SetAlgoLined
	\SetKwInOut{Input}{in}\SetKwInOut{Output}{out}
	\Input{CE sampling distribution $P(\cdot|\bm{\lambda}^*)$, initial sampling distribution parameter $\bm{\lambda}^*$, number of CE samples $n_{CE}$, number of elite samples $n_E$, number of observation history samples $n_{MC}$, maximum number of objective function evaluations $n_{max}$.}		
	\Output{optimal heuristic parameters $\bm{w^*}$, minimum total life-cycle cost $C^*$, surrogate cost function $f$}
	$l\leftarrow0$\;
	\While{$l\cdot n_{CE}<n_{max}$}{
		\For{$m\leftarrow1$ \KwTo $n_{CE}$}{
			$\bm{w}^{(m)}\sim P(\cdot|\bm{\lambda}^*)$\Comment*[r]{generate random heuristic parameter values}
			$cost\leftarrow 0$\;
			\For{$q\leftarrow1$ \KwTo $n_{MC}$}{
				generate an inspection and repair history $\mathbf{z}^{(q)}$ following strategy $\mathcal{S}_{\bm{w}^{(m)}}$\;
				$cost\leftarrow cost+\mathbf{E}_{\bm{\Theta}|\mathbf{z}_{1:n_T}^{(q)}} [C_{\text{tot}}|\bm{w}^{(m)},\mathbf{z}^{(q)}]$\Comment*[r]{Eq.~\ref{Eq:TotalCost} to \ref{Eq:RepCosts}}
			}
			$q_m\leftarrow cost/n_{MC}$\Comment*[r]{expected total life-cycle cost (Eq.~\ref{Eq:MCS})}
		}
		Store $\bm{w}^{(m)}$, $q_m$\;
		$\widehat{\bm{w}}^{(1)},..\widehat{\bm{w}}^{(n_{CE})}\leftarrow$Sort $(\bm{w}^{(1)},..\bm{w}^{(n_{CE})})$ in increasing order of $q_m$\;
		$\bm{\lambda}^* \leftarrow \arg\max_{\bm{\lambda}} \sum_{m=1}^{n_E} {\log P(\widehat{\bm{w}}^{(m)}|\bm{\lambda})}$\Comment*[r]{update $\bm{\lambda}^*$ with the elite samples}
		$l\leftarrow l+1$\;
	}
	$h\leftarrow GPR(\bm{w}^{(m)},q_{m})$\Comment*[r]{build the surrogate function}
	$\bm{w}^*\leftarrow \arg \min h$\;
	$C^*\leftarrow \min h$\;
	\Return{$\bm{w}^*,~C^*,~h$}
	\caption{Pseudo-code for the stochastic optimization}\label{Alg:2}
\end{algorithm}

\subsection{Summary}
\begin{figure}[!t]
	\centering
	\includegraphics[scale=0.6]{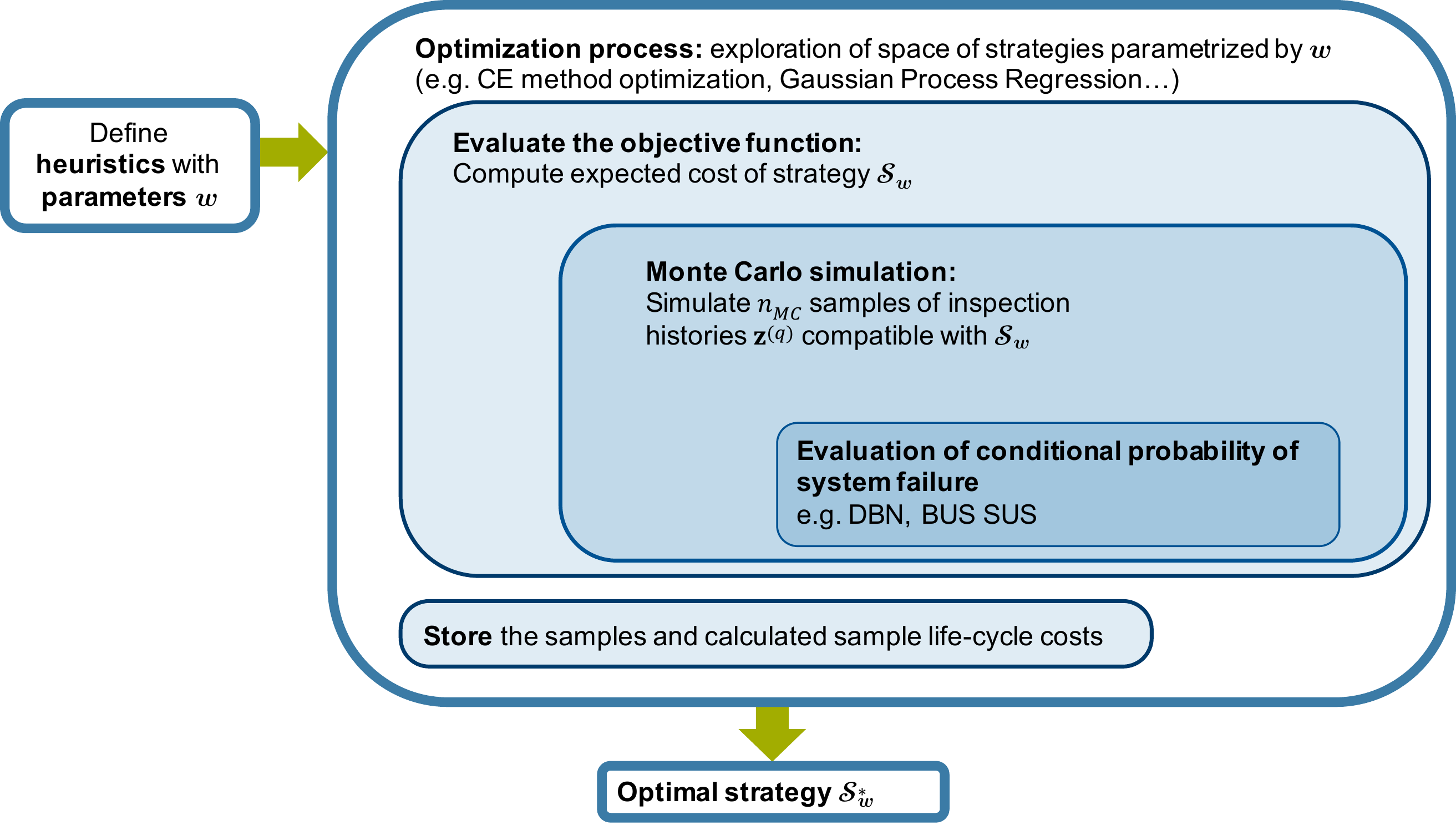}
	\caption{Steps of optimal  heuristic planning.}\label{Fig:Flowchart_DPS}
\end{figure}

The methodology is summarized in Figure~\ref{Fig:Flowchart_DPS}. It is applicable to any deteriorating system with multiple components. The methods for the optimization or for the evaluation of the expected cost of a strategy can be chosen freely, as they will not affect the underlying principles of the methodology.
For instance, the CE method can be upgraded to include optimal computing budget allocation (OCBA) \citep{He_et_al_10,Chen_LooHay_11}, which identifies the strategies for which $n_{MC}$ should be increased, to increase the confidence in the selection of the elite parameters. The computational effort is thereby optimized, as is the efficiency of the chosen optimization method.

\section{Heuristics for I\&M planning in structural systems}\label{Section:Heuristic}
The proposed approach requires the definition of appropriate heuristics. The heuristic should be sufficiently flexible to ensure that a sufficiently high number of strategies can be explored and that it can define a near-optimal strategy.

As an example, an overly simplistic heuristic is to never inspect and repair systematically at fixed time intervals $\Delta T$. Finding the solution of Equation \ref{Eq:Heuristic} is then equivalent to an optimization over one heuristic parameter $\bm{w}={\Delta T\in{\{1,..,T\}}}$. It is very likely that a heuristic strategy, which conditions the repair decision on observation outcomes, has a lower expected total life-cycle cost. 

\subsection{Choosing I\&M heuristics}

For I\&M planning of a multi-component deteriorating system, the possible decisions must answer the elementary questions of ‘when’, ‘where’, ‘what’ and ‘how’ to inspect, and ‘when’, ‘where’, ‘what’ and ‘how’ to repair. 

The questions ‘what’ and ‘how’ to inspect and repair are dependent on the deterioration mechanism and are in general determined by expert knowledge of the system, and the availability of repair and inspection techniques. They are not within the scope of this study and are not further discussed.

The questions ‘when’ and ‘where’ to repair can be answered in different ways; in corrective maintenance, repair or replacement occurs upon failure of the system; in systematic preventive maintenance, repair or replacement occurs at times fixed in advance, irrespective of any observation on the state of the system \citep{Barlow_Hunter_60,Nielsen_Sorensen_10}. The heuristic strategies investigated in this paper are relevant for a condition-based preventive maintenance,  where repair actions are decided based on the current condition of the system using inspection and monitoring data. In this approach, a component repair is triggered by an inspection result. With some generality, this repair decision can be parametrized by a threshold $D_{rep}$: Whenever the identified defect exceeds $D_{rep}$, a repair is triggered. 

 The inspection-related questions are more complex to parametrize and are discussed in the following paragraphs. They are treated sequentially, choosing first `when' and then `where' to inspect. 
 
 \subsection{`When' to inspect}
 
 An inspection can be planned to take place at specific times during the service life of the structure. It can also be triggered by exceeding a threshold $p_{th}$, for instance on the failure rate. This rate can usually be approximated by the annual probability of failure due to the high reliability of the structure. When a monitoring system is in place, an inspection might be triggered by the exceedance of a monitoring data threshold. 
 
 A number of scientific studies have focused on I\&M planning for a single component, answering the question `when' to inspect \citep[e.g.][]{Thoft_Sorensen_87,Grall_et_al_02,Straub_Faber_06,Nielsen_Sorensen_10}. By prescribing regular inspection times, or triggering inspections with a fixed threshold on the probability of failure of the component \citep{Faber_et_al_00}, these single-parameter heuristics have proven to yield a strategy that performs similarly in terms of cost and prescribed times of inspection as an optimal strategy found through POMDP or LIMID methods \citep{Nielsen_Sorensen_10a,Luque_Straub_13}. 
 
The choice of a heuristic parameter prescribing an inspection campaign might be based on operational constraints, for instance inspections might have to be planned at regular time intervals $\Delta T$. In other instances, they may arise from established practices. 
 
 \subsection{`Where' to inspect: component prioritization and value of information}\label{Sect:VoI}

 This question is specific to multi-component systems, where inspecting the entire system at every inspection campaign is suboptimal or not even feasible. Rather, one would typically inspect only a subset of all components during an inspection campaign. In order to identify the best components to inspect, we propose a heuristic prioritization rule described by two parameters.
 
 The first heuristic parameter is the number of components $n_I$ to inspect at each inspection campaign. A second heuristic parameter $\eta$ is introduced to prioritize the components for inspection at each campaign, as described in the following.
 
Because of the correlation among the component deterioration, inspecting one component can provide information about the condition of other components. 
The value of this information can in theory be quantified, as demonstrated in \citep{Straub_Faber_05}. 
It seems reasonable to perform the component inspections with the highest value of information (VoI). However, the calculation of the VoI is computationally challenging for multi-component systems.  

We introduce instead a Prioritization Index, $PI_k$,  to serve as a proxy for the VoI and prioritize the components for inspection.  This index is calculated for each component $k$ whenever an inspection campaign is launched, based on the information $\bm{Z}_{1:{i-1}}$ collected up to the last inspection time. 

The $PI_k$ considers two fundamental contributions to the VoI: (a) the reduction of the uncertainty on the condition of the inspected component and the corresponding effect on the system reliability; and (b) the reduction of the uncertainty on the condition of other components, through the components' interdependence. 

These reductions are related to two quantities. The first is the probability of failure of component $k$, $\text{Pr}(F_{c_k}|\bm{Z}_{1:{i-1}})$, conditional on all components' inspection outcomes up to that point. RBI planning revealed in particular that inspecting components with a higher probability of failure provides more information on the deterioration of other components than inspecting components with a lower probability of failure \citep{Straub_Faber_05}. 

The second quantity is the Single Element Importance measure for component $k$ ($SEI_k$), defined as the difference between the probability of failure of the intact system and the probability of failure of the system when only component $k$ has failed \citep{Straub_DerKiureghian_11}: 
\begin{equation} \label{Eq:SEI}
SEI_k= \text{Pr}(F_{s}|\overline{F_{c_{1}}},...,\overline{F_{c_{k-1}}},F_{c_k},\overline{F_{c_{k+1}}},...,\overline{F_{c_{N}}})-\text{Pr}(F_{s}|\overline{F_{c_{1}}},...,\overline{F_{c_{N}}})
\end{equation}

The VoI of inspection component $k$ is linked to the reduction of the probability of system failure. This probability can be expressed approximately as a linear function of $SEI_k$ and $\text{Pr}(F_{c_k}|\bm{Z}_{1:{i-1}})$ \citep{Bismut_et_al_17}, as detailed in \ref{App:PI}. Hence, the VoI increases with increasing $(SEI_k)$ and increasing $\text{Pr}(F_{c_k}|\bm{Z}_{1:{i-1}})$.

With an adjustable exponent $\eta$ that serves as a heuristic parameter, the Prioritization Index $PI_k$ combines the two effects described above,
\begin{equation}\label{Eq:PI}
PI_k=(SEI_k)^{\eta}\cdot \text{Pr}(F_{c_k}|\bm{Z}_{1:{i-1}}), \text{ with } \eta\geq 0.
\end{equation}

Other factors should be considered for the prioritization, such as the effect of varying component correlations, inspection quality, and cost of inspection. For instance, an underwater part of an offshore structure is more difficult and costly to inspect with the same accuracy as a part of the superstructure. Here, we limit the study to equi-correlated components, with equal inspection quality and cost. Additional factors with corresponding exponents (i.e. additional heuristic parameters) could however be introduced into Equation \ref{Eq:PI}.

\section{Numerical application: structural frame subject to fatigue}
\label{Section:Numerical}
\subsection{Structural model}\label{Section:Model_Descr}
 There are no standard procedures for I\&M planning of structures subject to fatigue deterioration. Time-based maintenance is in general not appropriate for this type of deterioration \citep{deJonge_et_al_17}. Condition-based maintenance is typically implemented by fixing an inspection interval between inspection campaigns, and inspecting all or selected components, following unknown heuristics. 

\begin{figure}[h!]
	\centering
	\includegraphics[scale=0.6]{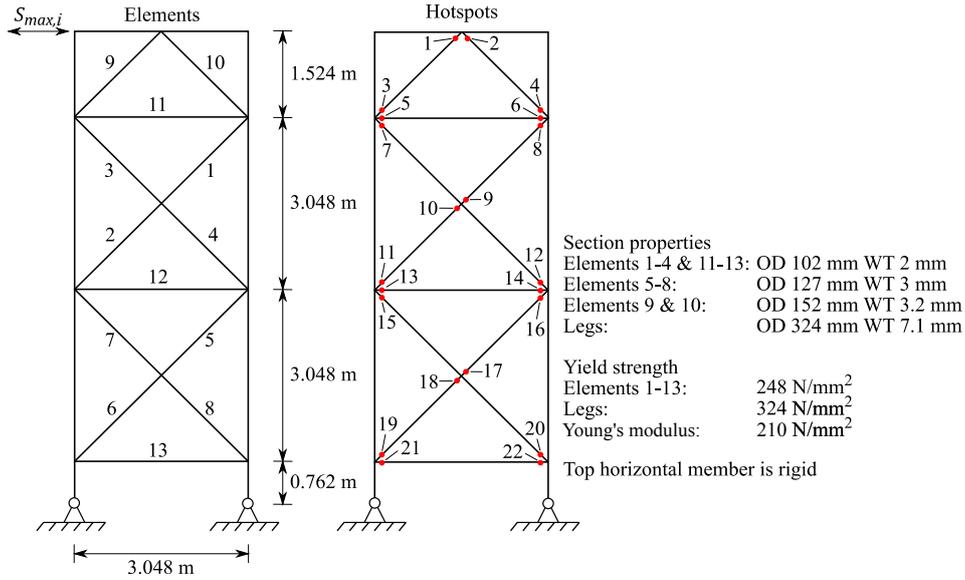}
	\caption{Zayas frame. The red dots indicate the locations of 22 fatigue hotspots. The frame is loaded laterally with yearly amplitude $S_{max,i}$. After \cite{Schneider_et_al_17}}
	\label{Fig:Zayas}
\end{figure}

We apply the described methodology to the I\&M planning of a steel structure representing a jacket support structure of an offshore wind turbine, which is typically subject to fatigue deterioration. We adopt the Zayas frame model \citep{Zayas_80}, often used for benchmark studies \citep[e.g.][]{Chen_Wai-Fah_88,Luque_Straub_16,Schneider_et_al_17,Schneider_19}. This steel frame is composed of two vertical legs, braced by 13 tubular members; 22 fatigue hotspots located at the welds are identified as per Figure~\ref{Fig:Zayas} and constitute the model components. The states of deterioration of the components are correlated (see Section~\ref{Section:Deterioration_Model}). The frame is loaded laterally with a time-varying load, $S_{max,i}$, which represents the maximum load occurring within one time step, or year. We assume that the maximum annual loads are independently distributed, following a lognormal distribution, with mean $50$kN and coefficient of variation $0.53$. Failure is determined by the system components deterioration states and the response of the damage structure to the applied load. The pushover analysis for the frame is from \cite{Schneider_et_al_17}. The ultimate resistance of the undamaged frame is $282$kN.

\subsection{Deterioration model}\label{Section:Deterioration_Model}
For each component $k$, $D_{k,i}$ indicates the crack depth at time $t_i$. The transition from $D_{k,i}$ to $D_{k,i+1}$ is obtained from Equation~\ref{Eq:Paris}, which models the fatigue crack growth by Paris’ law,
\begin{equation} \label{Eq:Paris}
\frac{dD}{dt}=C [\Delta S_{e}^{M} \pi^{\frac{M}{2}}] \cdot D(t)^{\frac{M}{2}},
\end{equation}
where $C$ and $M$ are empirical material parameters. They are correlated with a correlation coefficient close to $-1$, hence $C$ is here expressed as a linear function of $M$ \citep{Straub_04}. The initial crack depth is noted $D_{k,0}$ and failure of a component is defined as the fatigue crack size exceeding a critical depth $d_{cr}$. 

The fatigue stress range $\Delta S$ is described by a Weibull distribution with scale and shape parameters $K$ and $\lambda$. The distribution of the equivalent fatigue stress range $\Delta S_{e}=(\mathbf{E}[\Delta S^{M}])^{\frac{1}{M}}$ is defined by Equation~\ref{Eq:Weibull} \citep{Straub_04}, as
\begin{equation} \label{Eq:Weibull}
\Delta S_{e}=K\cdot\Gamma \left(1+ \frac{M}{\lambda}\right)^{\frac{1}{M}}.
\end{equation}
$K$ and $M$ are unknown parameters and are assumed constant during the deterioration process. In order to preserve the Markovian assumption, the random variables $K_{k,0\leq i\leq n_T}$ and $M_{k,0\leq i\leq n_T}$ are introduced in the Bayesian network, such that $K_{k,i+1}=K_{k,i}=K_{k}$ and $M_{k,i+1}=M_{k,i}=M_{k}$. The corresponding DBN is shown in Figure~\ref{Fig:DBN}.

The distribution of $K_k$ for each component $k$ is lognormal, and the mean values are calibrated to a chosen fatigue design factor (FDF). The FDFs represent the ratio between the fatigue life of a component and the service life of the structure. The calibration procedure is explained in \ref{App:CalibrationFatigue}. Table~\ref{Tab:K_FDF} lists the components, their FDFs, and the corresponding means of $K_{k}$. 

Additionally, the initial crack depths of all components $D_{k,0}$ are equi-correlated with factor $\rho_{D_0}$, as are the stress and material parameters, $K_{k}$ and $M_{k}$ with factors $\rho_{K}$ and $\rho_{M}$ respectively. The correlation is reflected in the computation of the probability tables conditional on the hyperparameters $\bm{\alpha}$ of the DBN, following the procedure of \citep{Luque_Straub_16}.

The model parameters and correlations are summarized in \ref{App:ModelParam}.

\subsection{Inspection and repair model}\label{Section:Insp_model}
  
The observation outcome ${Z}_{k,i}$ is a random variable defined conditionally on $D_t$. ${Z}_{k,i}$ can take the value  \{$Z_{k,i}=0$\}, indicating \{\textit{`no crack detected'}\}, or values larger than $0$, reflecting a measured crack size. Here, the probability of detection (PoD) $\text{Pr}(Z_{k,i}>0|D_t=d)$ is:
\begin{equation} \label{Eq:PoD}
PoD(d)=1-\text{exp}\left(-\frac{d}{\xi}\right)
\end{equation}

In case of detection, the measurement ${Z}_{k,i}$ is normally distributed with mean $D_{k,i}$ and standard error $\sigma_\epsilon$. The corresponding hybrid distribution is defined in Equation~\ref{Eq:inspection}, where $\varphi(.)$ is the standard normal probability density function, and $\Phi(.)$ is the standard normal cumulative distribution function:
\begin{equation} \label{Eq:inspection}
\begin{cases}
\text{Pr}(Z_{k,i}=0|D_{k,i}=d)=1- PoD(d)\\
f_{Z_{k,i}|D_{k,i}=d}(Z_{k,i}=z)=PoD(d)\cdot \frac{1}{1-\Phi \left(\frac{-d}{\sigma_\epsilon}\right)}\cdot \varphi\left( \frac{z-d}{\sigma_\epsilon}\right) & \text{if } z>0.
\end{cases}
\end{equation}

If the damage exceeds a threshold $D_{rep}$, the component is repaired immediately and completely. This corresponds to restoring it to its initial state, i.e. the probability distribution of the damage after a repair is the one of $D_{k,0}$. The repaired state is assumed to be uncorrelated with the initial state.

Following from DBN model structure described in Section~\ref{Subsection:DBN} and in Figure~\ref{Fig:DBN}, the inspection outcome of one component affects the posterior distribution of the crack size of all other components, whether or not the inspected component is repaired.

\subsection{Costs of I\&M actions}
  
I\&M actions incur mobilization costs, repair costs, costs for material supply, workmanship, inspection techniques, and sometimes downtime. 
Here, all components are attributed the same cost of inspection and repair, which are constant over time. 
This choice simplifies the problem specification; alternative cost models do not affect the computation efforts or the accuracy of the method.

The following costs are considered:
\begin{itemize} [nosep] 
	\item $c_C$: cost of launching an inspection campaign. It includes the cost of transporting inspection operators to site, and potentially the cost of impairing the operation of the system.
	\item $c_I$: cost of inspection per component. It accounts for the time spent to inspect one component during an inspection campaign.
	\item $c_R$: cost of repairing one component. It also includes the associated downtime.
	\item $c_F$: cost incurred if the structure fails, including full replacement costs and life-cycle costs of the new structure.
\end{itemize}
All costs are discounted to their present value by the discount factor $\gamma(t)=\frac{1}{(1+r)^t}$, where $r$ is the annually compounded discount rate. The parameters of the cost function and discount factor are provided in Table \ref{Tab:Cost}. 

\begin{table}[htbp] 
	\centering
	\caption{Cost parameters\label{Tab:Cost}}
	\begin{tabular}{c |c c c c c}
		
		Parameter & $c_C$ & $c_I$ &$c_R$ & $c_F$ & $r$\\
		\hline
		Value & $1$ &  $0.1$ & $0.3$ & $3\cdot 10^3$ & $0.02$\\
		
	\end{tabular}
\end{table}

\subsection{Heuristic parameter choice}\label{Section:Param_Choice}
The investigated heuristic includes the following parameters $\bm{w}=\{\Delta T, p_{th}, n_I, \eta\}$:
\begin{enumerate}[label=(\alph*)]
	\item Inspection campaigns are carried out at fixed inspection intervals $\Delta T\in\{1,2,…40\}$ [years].
	\item Additionally, when the annual probability of failure of the system exceeds the probability threshold $p_{th}\in[0,1]$, an inspection campaign is carried out. 
	\item The number of hotspots to be inspected at every inspection campaign is $n_I\in\{1,2,…,22\}$.
	\item The hotspots are prioritized for inspection following Section~\ref{Sect:VoI} with parameter $\eta\geq0$.
\end{enumerate}
The repair threshold $D_{rep}\geq0$ can also be included as a heuristic parameter. For the purpose of this study, a component is repaired if the deterioration exceeds a threshold value $D_{rep}=0$ at the inspection. This implies that every identified defect is repaired. The results documented in Section~\ref{Section:Results} show that the expected repair costs are low even with this choice, which indicates that optimizing $D_{rep}$ would not significantly affect the expected life-cycle cost.

\subsection{Optimization set up and computation}
 
The CE sampling algorithm is run with MATLAB on a 2.6GHz computer with 24 quad-core processors. The algorithm is optimized for the computation of the conditional probability of system failure. One history life-cycle cost is computed in about 20 CPU minutes, but these computations can be run in parallel.

For the CE optimization method of Algorithm~\ref{Alg:2}, lognormal sampling distributions are employed for the heuristic parameters $p_{th}$ and $\eta$. 
For the discrete parameters $\Delta T$ and $n_I$, a truncated normal distribution is selected, so that the sampled values are within the acceptable bounds; the sampled values are then rounded to the nearest integer.
In this numerical application, all sampling distributions are kept uncorrelated and the heuristic parameters are optimized one after the other in a recursive manner. This was mainly for ease of graphical representation of the results. Correlating the distributions might lead to more efficient sampling during the optimization and faster convergence towards the optimal parameter values.
 
We fix $n_{MC}=1$. About 1600 samples of total life-cycle costs for different strategies are drawn as per Equation \ref{Eq:TotalCost}. To obtain a surrogate of the total life-cycle costs, a GPR is performed on the logarithm of the sample costs. This guarantees that the surrogate function stays positive in the original space. 
The GPR is implemented with the 
MATLAB2018 function \textit{fitrgp.m} and a squared-exponential kernel. The minimum of this surrogate cost function is calculated with the MATLAB2018 function \textit{surrogateopt.m}.

GPR can also be performed on all four cost components separately: inspection campaign, component inspection, repair and failure risk. However, it was found that the total cost predictions obtained by summing up the four resulting Gaussian processes are not as stable as the predictions obtained by the GPR on the total life-cycle cost, and are therefore not used to find the optimal heuristic parameter values. Figure~\ref{Fig:SampleCosts} compares the prediction of the expected total life-cycle cost with the two methods, for selected strategies.

We consider one adaptation of the I\&M plan after the first inspection campaign. For the adaptation, the deterioration model is updated with the inspection outcomes at the time prescribed by the initial strategy. Then, 1600 deterioration histories are simulated from the posterior distributions, using BUS SUS \citep{Straub_Papaioannou_15}. Based on these samples, total life-cycle costs are calculated for different strategies, and the strategy optimization is performed as outlined in Section~\ref{Section:Adaptive}.

\section{Results}
\label{Section:Results}

\subsection{Conditional annual risk computed with DBN}
\begin{figure}[t!]
	\centering
	\captionsetup[subfigure]{justification=centering}
	
	\begin{subfigure}[t]{0.5\columnwidth}
		
		\includegraphics[trim=0 20 0 0,clip,scale=0.4]{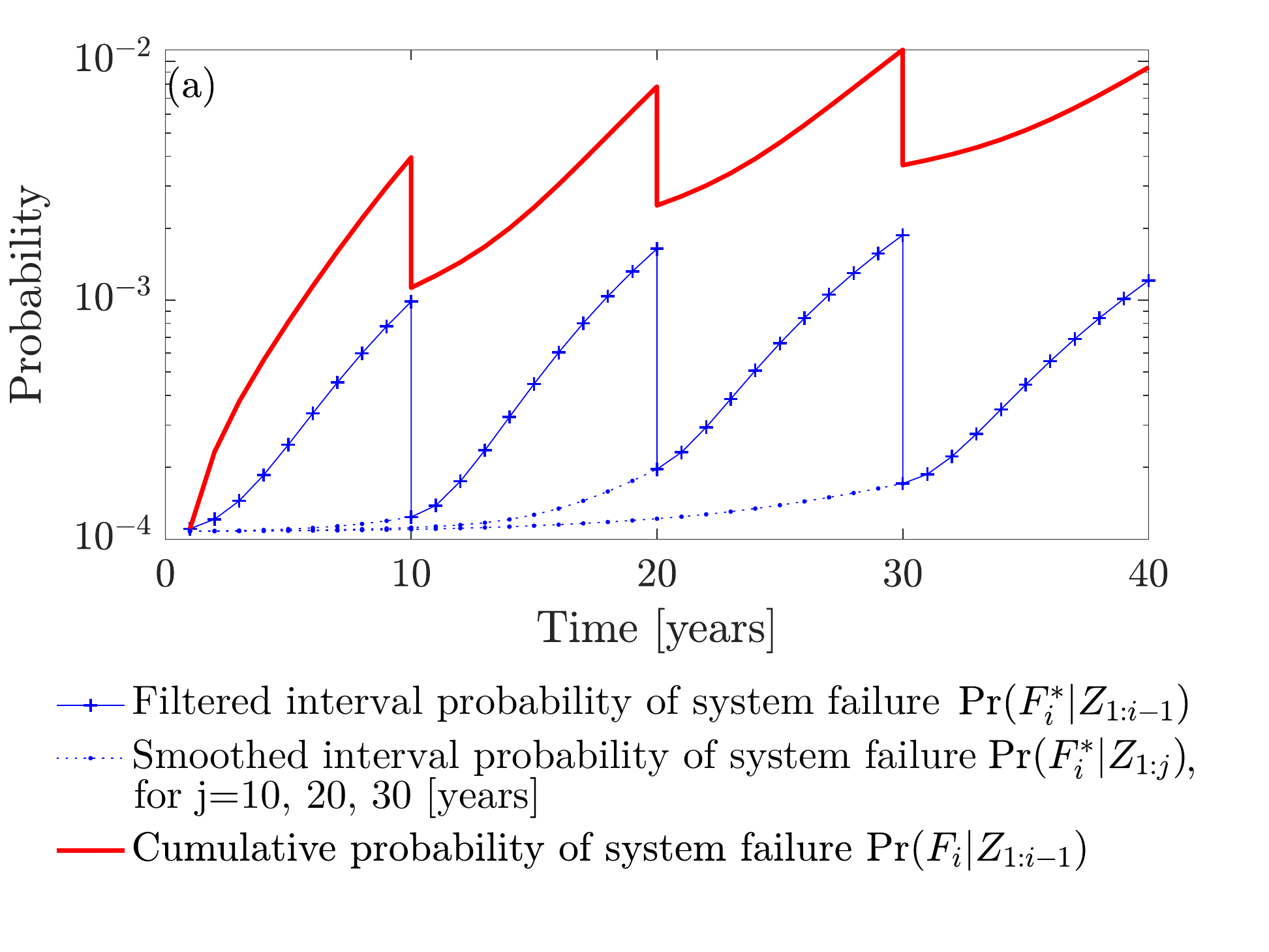}
		\caption{~}
		\label{Fig:SamplePoF}
	\end{subfigure}%
	\begin{subfigure}[t]{0.5\columnwidth}
		
		\includegraphics[trim=0 20 0 0,clip,scale=0.4]{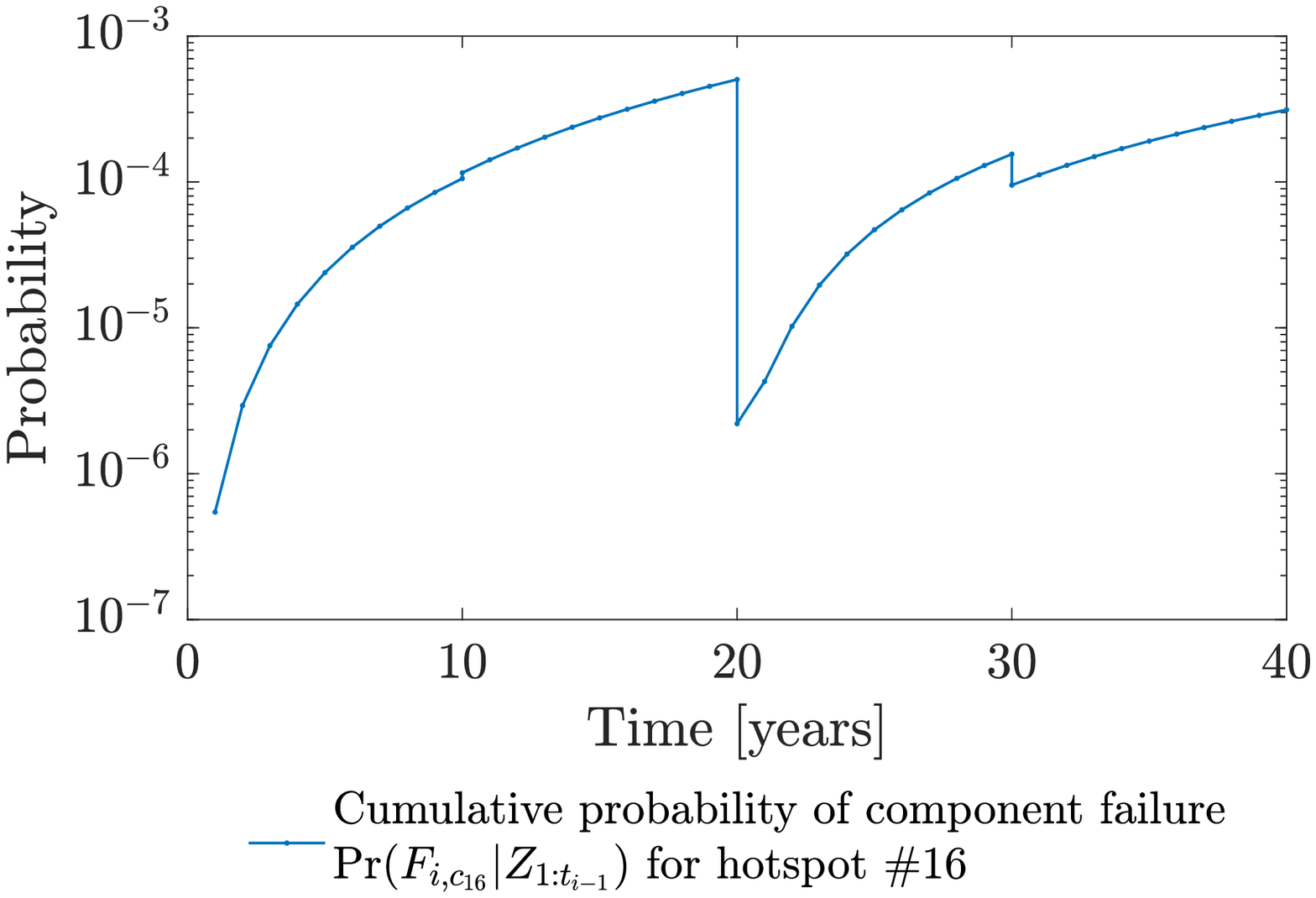}
		\caption{~}
		\label{Fig:ComponentPoF}
	\end{subfigure}

	\caption{(a) Evolution of the interval probability and cumulative probability of system failure for a sample observation history following strategy $\mathcal{S}_{\bm{w}}$, with ${\bm{w}}=\{\Delta T=10,~p_{th}=1\cdot 10^{-2},~n_I=4,~\eta=1\}$. The discontinuities indicate that an inspection is performed and the reliability is updated. The dotted blue lines represent the smoothed probabilities $\text{Pr}(F_j^*|\bm{Z}_{1:i})$, for $j\leq i$, updated with the current system information; (b) corresponding evolution of the probability of failure for hotspot $\sharp16$. The discontinuities at years \{10, 30\} are due to inspections on other components in the system. Hotspot $\sharp16$ is inspected at year 20, and is never repaired in this simulation.}
	\label{Fig:Sample_PoF}
\end{figure}

 Figure~\ref{Fig:SamplePoF} depicts an example of the evolution of the computed probabilities for a sample observation history, following the strategy described by heuristic parameters $\{\Delta T=10,~p_{th}=1\cdot 10^{-2},~n_I=4,~\eta=1\}$. %sample history #92
The corresponding probability of failure of a selected hotspot is plotted in Figure~\ref{Fig:ComponentPoF}. %hotspot #12 in Matlab
It should be noted that the discretization scheme influences the accuracy of these computations, notably when performing Bayesian smoothing. The associated error is investigated in \citep{Zhu_Collette_15} and is mainly due to the rough discretization in the failure domain of the component. However, we have found that it does not significantly impact the evaluation of $\text{Pr}(F_i^*| \bm{w},\bm{Z})$.

  \subsection{Optimal heuristic strategy at time $t=0$}

\begin{figure}[!ht]
	\centering
	\includegraphics[scale=0.6, trim=40 0 30 0]{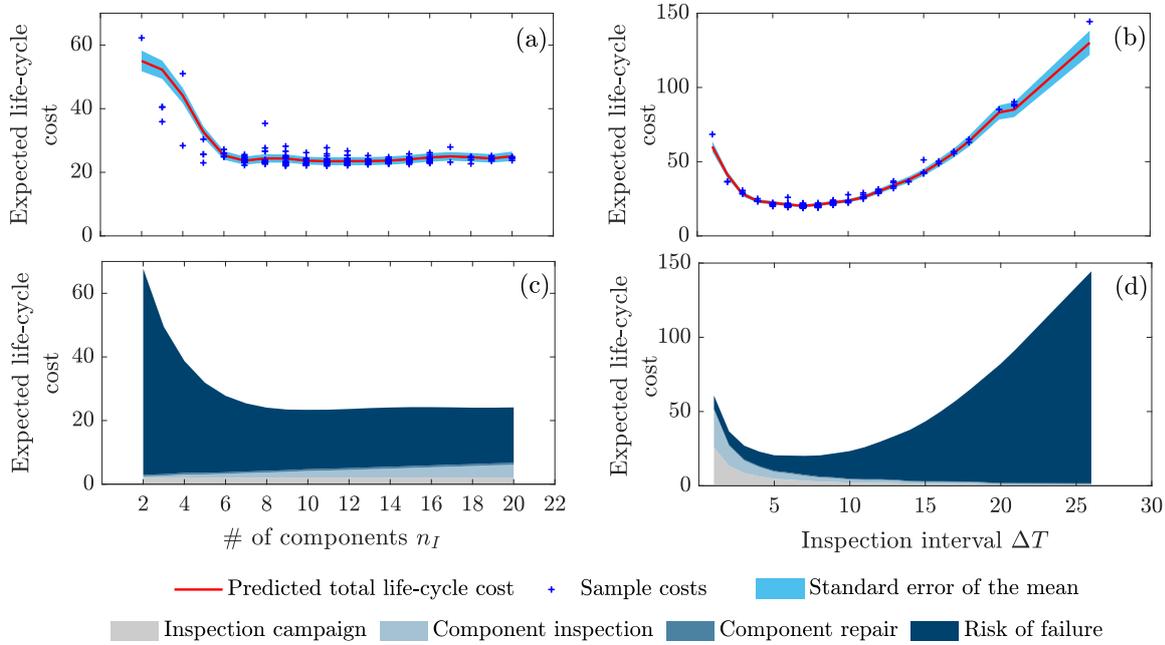}
	\caption{Sample total life-cycle cost with corresponding estimated mean value and standard deviation of the prediction (a-b). The prediction for the cost breakdown is shown in (c-d). There is a slight difference in the prediction of the total cost between (a-b) and (c-d), which comes from the GPR. (a) and (c): varying $n_I$ with fixed values \{$\Delta T=10$, $p_{th}=1\cdot 10^{-2}$, $\eta=1$\}; (b) and (d): varying $\Delta T$ with fixed values \{$p_{th}=1\cdot 10^{-2}$, $n_I=10$, $\eta=1$\}. }
	\label{Fig:SampleCosts}
\end{figure}

First, we investigate how the expected cost varies in function of the different heuristic parameters. Figure~\ref{Fig:SampleCosts} shows the total life-cycle cost in function of $n_I$ and $\Delta T$ for sample strategies from the CE method. The GPR surrogate estimate is also shown.

\begin{figure}[!ht]
	\centering
	\includegraphics[width=1\linewidth, trim=0 10 30 0]{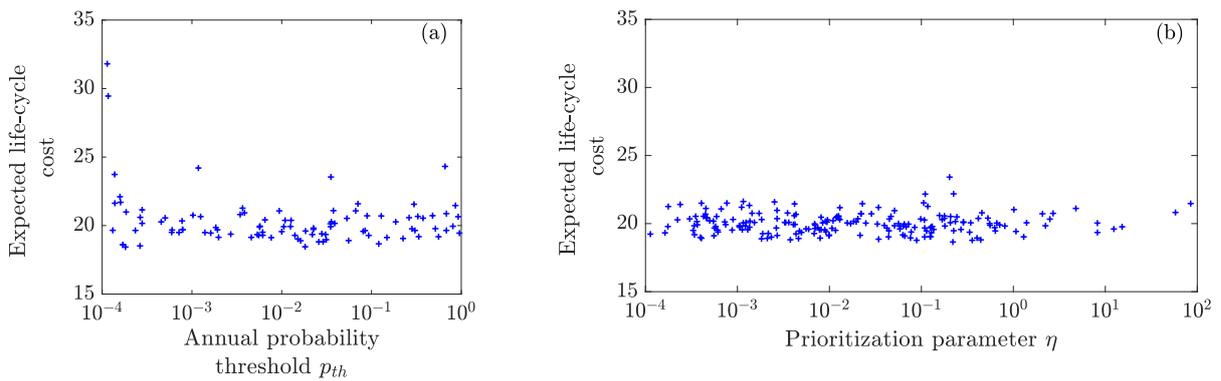}
	\caption{Sample total life-cycle costs, for varying probability threshold $p_{th}$ and prioritization parameter $\eta$. The risk of failure decreases with the probability threshold. However for some values of $p_{th}$ (here lower than $10^{-4}$), the inspection and repair costs needed to satisfy the threshold requirement increase strongly. Probability thresholds that are too small to meet are assigned an arbitrary cost of 1000. (a): $\Delta T=7$, $n_I=9$, $\eta=1.3$ (b): $\Delta T=7$, $n_I=10$, $p_{th}=5.2\cdot 10^{-3}$.}\label{Fig:SampleCosts_1}
\end{figure}

As expected, the total life-cycle risk of failure decreases with increasing number $n_I$ of hotspots inspected at every inspection campaign. Similarly, strategies with small inspection intervals $\Delta T$ are costly because of the frequency of the inspections; in contrast, strategies with large $\Delta T$ suffer from a larger uncertainty and thus an increased risk of failure. Samples drawn in a similar way for varying $p_{th}$ and $\eta$ show a less clear trend, as can be seen in Figure~\ref{Fig:SampleCosts_1}.

The minimum of the (continuous) surrogate cost function is achieved for the values  $\{ \Delta T=7.0,~p_{th}=1.9\cdot10^{-3},~n_I= 9.4,~ \eta=1.3\}$. The values for $\Delta T$ and $n_I$ are rounded to the nearest integer. 

\begin{figure}[t!]
		\centering
	\captionsetup[subfigure]{justification=centering}
	\begin{subfigure}[t]{0.5\columnwidth}
	\includegraphics[trim=15 0 25 0,clip, scale=0.45]{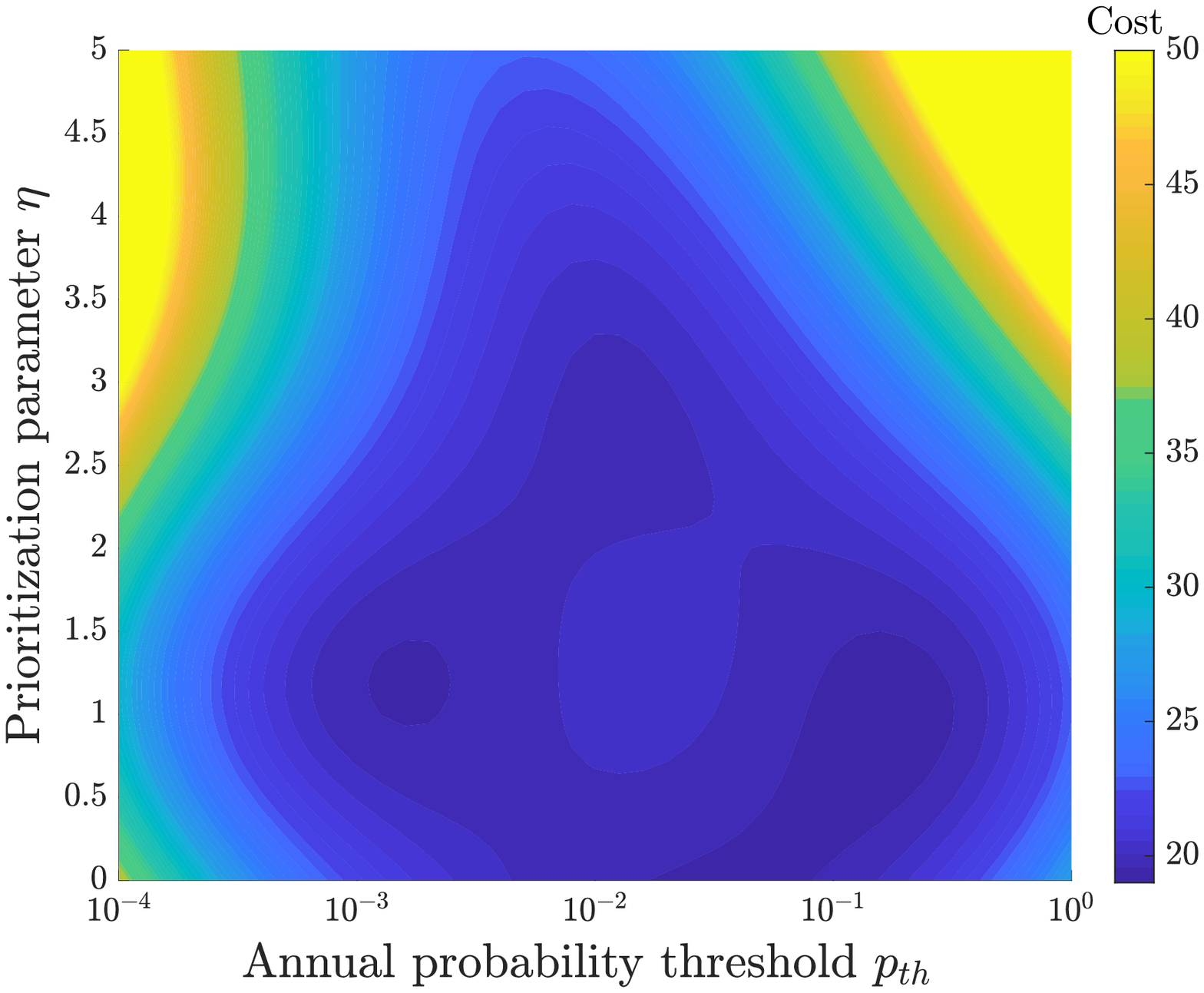}
	\caption{~}\label{Fig:SurrogateOpt_2}
	\end{subfigure}%
	\begin{subfigure}[t]{0.5\columnwidth}
		\includegraphics[trim=20 0 25 0,clip, scale=0.45]{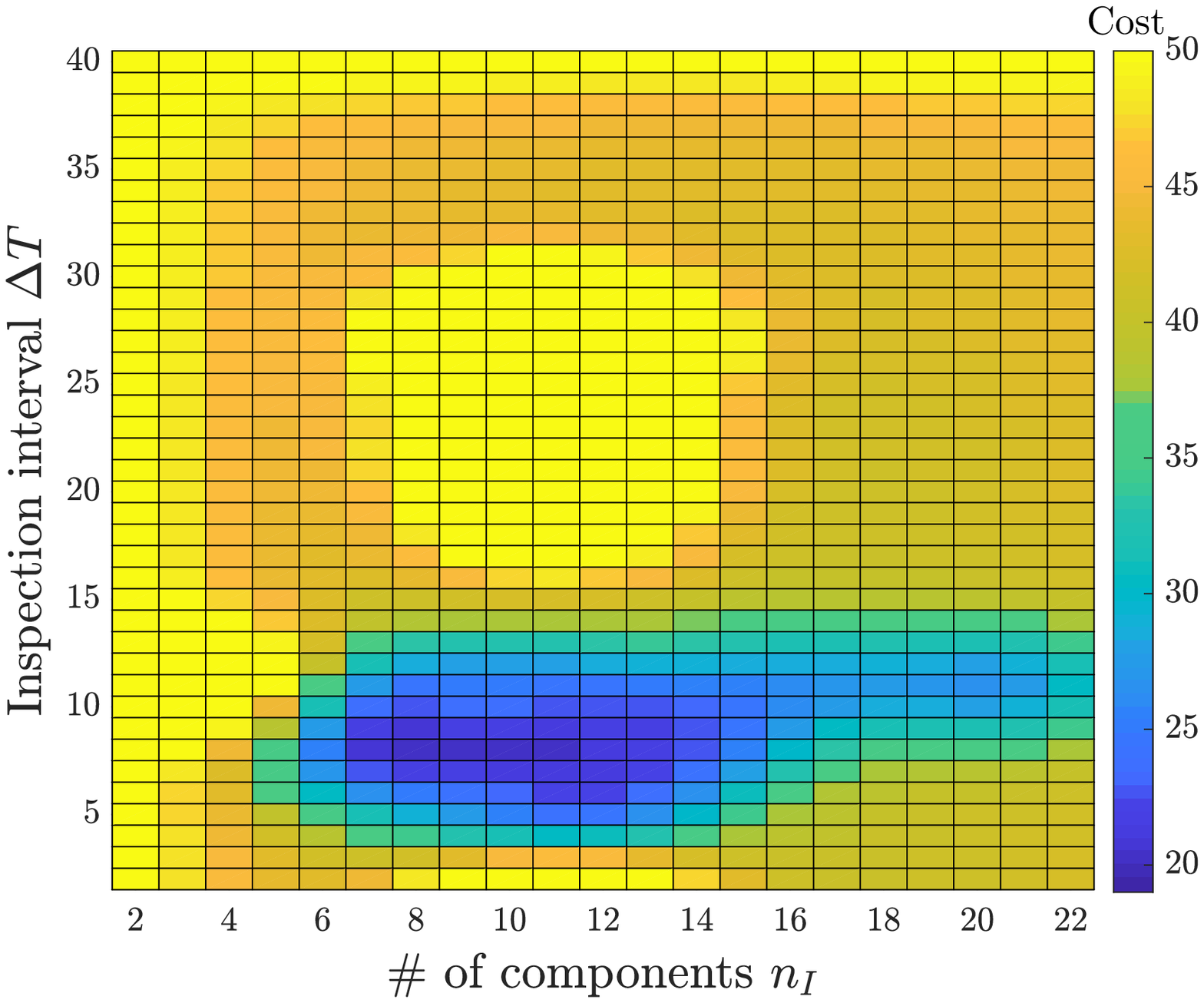}
		\caption{~}\label{Fig:SurrogateOpt_1}
\end{subfigure}
\caption{(a): Surrogate cost function around the optimum for the prior optimization at $t=0$ for varying heuristic parameters $p_{th}$ (logarithmic scale) and $\eta$, with $\Delta T=7$[years] and $n_I=9$. (b): Surrogate cost function around the optimum for the prior optimization at $t=0$ for varying (integer) heuristic parameters $n_I$ and $\Delta T$, with $p_{th}=2 \cdot 10^{-2}$ and $\eta=1.3$.}
\end{figure}

Figure~\ref{Fig:SurrogateOpt_2} depicts the contours of the GPR surrogate function, as a function of $p_{th}$ and $\eta$, for fixed $\Delta T=7$[years] and $n_I=9$. Within a large domain of parameters $p_{th}$ and $\eta$, the surrogate cost varies between $19$ and $21$. 
Hence the total life-cycle cost is not sensitive to $p_{th}$ and $\eta$ when the values $\Delta T=7$[years] and $n_I=9$ are set.

The GPR only provides an approximation of the underlying function; it reports a standard error of 5\% on the estimated expected life-cycle cost at the point minimizing the surrogate. To account for this error, we computed 200 additional Monte Carlo samples of deterioration histories and inspection outcomes at the point minimizing the surrogate, which yields an expected total life-cycle cost of $21.4$. We also evaluated 200 Monte Carlo samples for a second strategy, increasing the value of $p_{th}$ by an order of magnitude, i.e. $\{ \Delta T=7,~p_{th}=2\cdot10^{-2},~n_I= 9,~ \eta=1.3\}$. This is motivated by Figure~\ref{Fig:SurrogateOpt_2}. For this we obtain an expected total life-cycle cost of $20.1$. Both costs are similar, which confirms that the parameter $p_{th}$ does not considerably affect the expected life-cycle cost of a strategy. Nevertheless, we adopt the latter strategy as the best strategy, $\mathcal{S}_{\bm{w}_0^*}$, for the prior model, since it gives the smaller expected cost. We highlight the result in Table~\ref{Tab:Opt_0}, in which we also include the value given by the surrogate at that point.  The corresponding breakdown of the expected total life-cycle cost into its four components is found in Table~\ref{Tab:CostBreak_0}.

\begin{table}[h]
	\centering
	\caption{Parameters $\bm{w}_0^*$ and expected total life-cycle cost of the optimal strategy at time $t_0=0$.\label{Tab:Opt_0}}
	\begin{tabular}{c| c| c| c||c|c}
		$\Delta T$ & $p_{th}$ & $n_I$ & $\eta$ & $\mathbf{E}[C_{\text{tot}}|\bm{w}_0^*]_{MC}$& $\mathbf{E}[C_{\text{tot}}|\bm{w}_0^*]_{\text{surrogate}}$  \\
		\hline
		$7$ & $2 \cdot 10^{-2}$ & $9$ & $1.3$ & $20.1$ & $20.1$\\

	\end{tabular}
\end{table}

The Monte Carlo samples also give an estimation of the annual repartition of the cost. Figure~\ref{Fig:AnnualCosts} shows the expected costs associated with the optimal  I\&M plan. 
Notably, the annual risk increases in the years before the first inspection and is progressively reduced with each additional inspection campaign, until it reaches a stationary value.

\begin{table}[t!]
    \centering
    \caption{Breakdown of the expected total life-cycle cost for strategy $\mathcal{S}_{\bm{w}_0^*}$.}
    \begin{tabular}{c|c}
    Cost component   & Expected cost\\ 
    \hline
        System failure & $13.0$ \\
        Inspection campaign & $3.3$\\
        Component inspection & $3.0$\\
        Component repair & $0.8$\\
        \hline
    \end{tabular}
    \label{Tab:CostBreak_0}
\end{table}

  \subsection{Adaptive strategy at time $t_1=7$ years}
The adaptive case is now investigated. Following $\mathcal{S}_{\bm{w}_0^*}$, the first inspection campaign is carried out at year $t_1=7$, with  inspections of components  \{8,9,10,11,16,17,18,19,20\}, at a cost of $c_{ini}=9*c_I+c_C=1.9$.  
No damage is detected. This information is stored as $\mathbf{z}_7$. 

At this point, the operator decides to improve the strategy adaptively, rather than continue with strategy $\mathcal{S}_{\bm{w}_{0}^*}$.

The optimal strategy $\mathcal{S}_{{\bm{w}_1^*}_{|\mathbf{z}_7}}$ obtained for the posterior model is characterized by the heuristic parameter values in Table~\ref{Tab:Opt_1}, with associated expected total life-cycle cost $\mathbf{E}[C_{\text{tot}} |{\bm{w}_1^*}_{|\mathbf{z}_7}]_{\text{surrogate}}=15.9$. This value is discounted to time $t_1$, and excludes the initial cost $c_{ini}$.

\begin{table} [h!]
	\centering
	\caption{Parameters $\bm{w}_1^*$ and expected total life-cycle cost of the optimal strategy at time $t_1=7$, after no detection of damage on components \{8,9,10,11,16,17,18,19,20\}.\label{Tab:Opt_1}}
	\begin{tabular}{c| c| c| c ||c}
	$\Delta T$ & $p_{th}$ & $n_I$ & $\eta$ &$\mathbf{E}[C_{\text{tot}}|{\bm{w}_1^*}_{|\mathbf{z}_7}]$ \\
	\hline
		$9$ &  $3\cdot 10^{-2}$ & $9$ & $2.2$ & 15.9\\

	\end{tabular}
\end{table}

\begin{figure}[t!]
	\centering
	\captionsetup[subfigure]{justification=centering}
		\begin{subfigure}[t]{0.5\columnwidth}
		\includegraphics[trim=15 0 25 0, clip,scale=0.45]{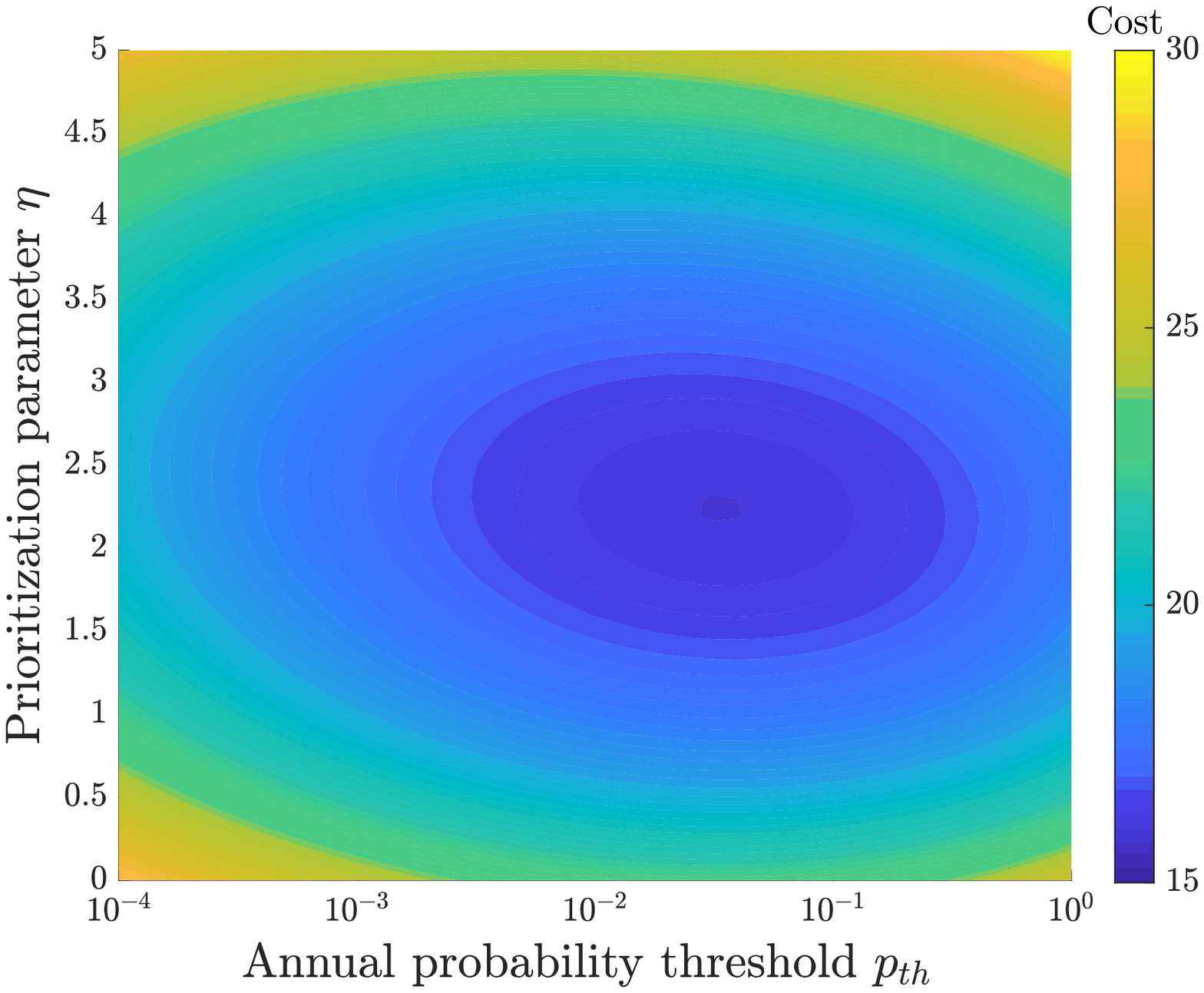}
		\caption{~}\label{Fig:SurrogateOpt_2_adapt}
	\end{subfigure}%
	\begin{subfigure}[t]{0.5\columnwidth}
		\includegraphics[trim=20 0 25 0, clip, scale=0.45]{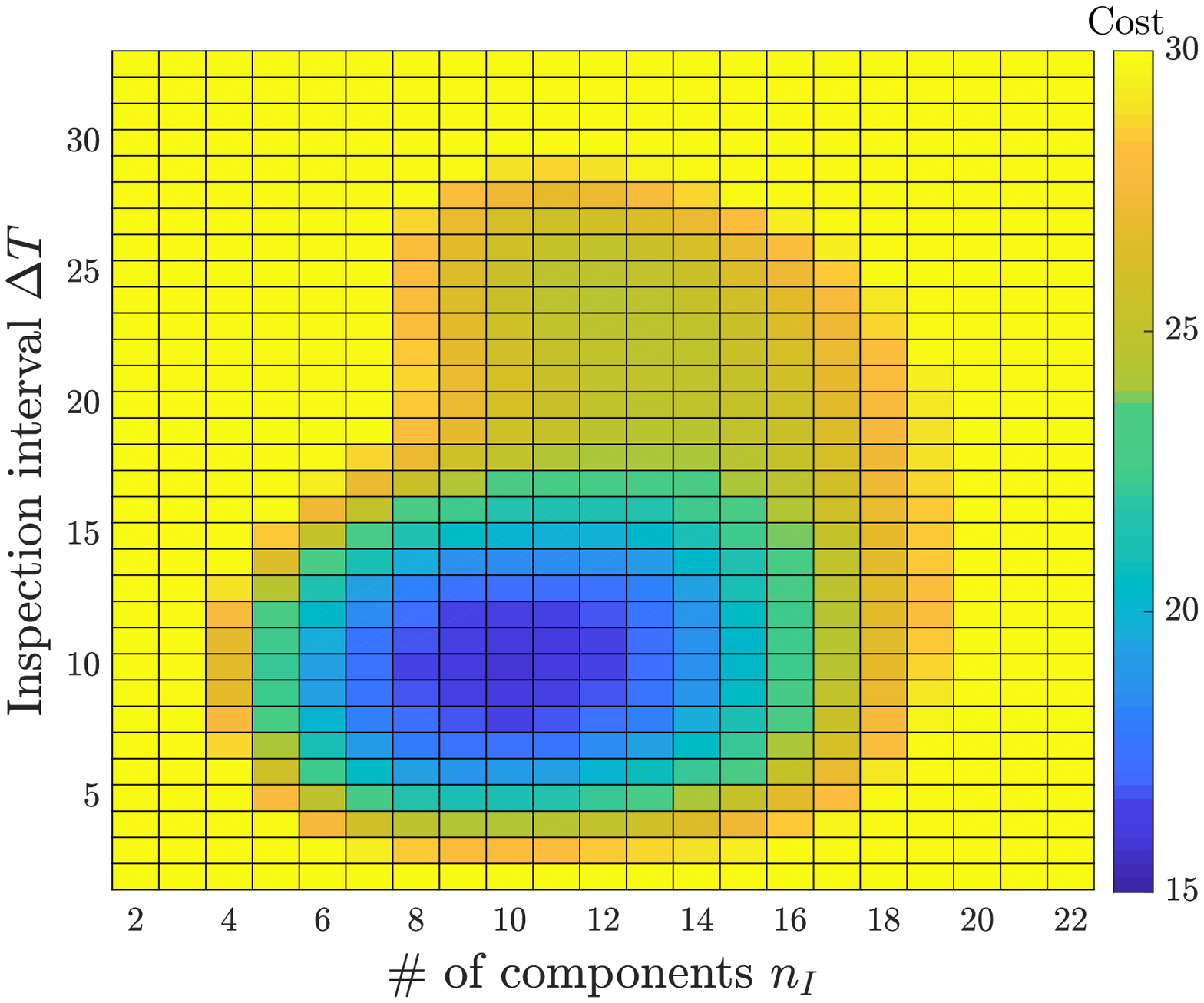}
		\caption{~}\label{Fig:SurrogateOpt_1_adapt}
	\end{subfigure}
	\caption{(a): Surrogate cost function around the optimum for the adaptive optimization at $t_1=7$ for varying heuristic parameters $p_{th}$ (logarithmic scale) and $\eta$, with $\Delta T=9$[years] and $n_I=9$. (b): Surrogate cost function around the optimum for the adaptive optimization at $t_1=7$ for varying (integer) heuristic parameters $n_I$ and $\Delta T$,  with $p_{th}=3 \cdot 10^{-2}$ and $\eta=2.2$.}
\end{figure}

Figures~\ref{Fig:SurrogateOpt_2_adapt} and~\ref{Fig:SurrogateOpt_1_adapt} depict the contours of the GPR surrogate cost function around the optimum values of the heuristic parameters at time $t_1$. 

The updated optimal heuristic parameters reflect the information gained at time $t_1$ and the reduction of uncertainty about the state of the structure. The optimal inspection interval increases to $\Delta T=9$, as does the optimal threshold on the annual probability of system failure, $p_{th}=3\cdot 10^{-2}$. The prescribed number of hotspots to be inspected remains at $n_I=9$. Furthermore, the updated prioritization component $\eta$ increases, meaning that it gives more weight to the components’ importance in the system; hence the components with a higher $SEI$ are more likely to be inspected several times during the service life.

\begin{figure}[t!]
	\centering
	\includegraphics[scale=0.5]{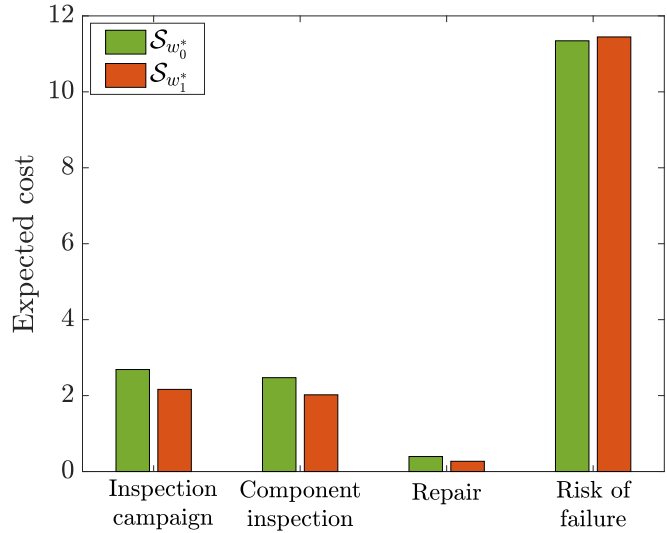}
	\caption{Comparison of expected cost for inspection, repair and failure for strategies $\mathcal{S}_{\bm{w}_0^*}$ and $\mathcal{S}_{\bm{w}_{1}^*}$, conditional on $\mathbf{z}_7$.}\label{Fig:CostCompare_2}
\end{figure}

In order to compare the initial strategy with the adapted strategy, the expected cost of $\mathcal{S}_{\bm{w}_0^*}$ conditional on $\mathbf{z}_7$ is evaluated with the surrogate cost function. We find that $\mathbf{E}[C_{\text{tot}} |\bm{w}_{0}^*,\mathbf{z}_7]_{\text{surrogate}}=16.9$, excluding the initial cost $c_{ini}$. The expected net gain of changing from strategy $\mathcal{S}_{\bm{w}_0^*}$ to strategy $\mathcal{S}_{\bm{w}_{1}^*}$, after detecting no damage at the first inspection is therefore
\begin{equation}
\mathbf{E}\left[C_{\text{tot}} |\bm{w}_{0}^*,\mathbf{z}_7\right]-\mathbf{E}[C_{tot}|{\bm{w}_1^*}_{|\mathbf{z}_7}]=1.0.
\end{equation}
Figure~\ref{Fig:CostCompare_2} shows that this gain comes from decreasing the number of inspections and repairs, which can be achieved without significantly increasing the risk. 

\section{Concluding remarks}
\label{Section:ConcludingRemarks}
In this paper, we present an integral framework for planning inspections and maintenance actions in structural systems. The framework, which draws upon our earlier works \citep{Bismut_et_al_17,Luque_Straub_17}, for the first time enables such a planning for a general structural system as a whole. It explicitly models deterioration at each component and includes the dependencies among multiple components. It accounts for the interaction among the components within the structural system and their effect on the system reliability. The framework also propagates the information collected at individual component inspections to the rest of the system. 

The framework is based on the use of heuristics for prescribing inspection and maintenance strategies. This heuristic approach only results in an approximately optimal strategy. When new information is available, an improved heuristic may be found, which is more optimal under the a-posteriori model. This motivated us to propose an adaptive approach, in which a new optimal heuristic is identified after inspection results are obtained. As we prove in Section \ref{Section:Adaptive}, this adapted strategy is always at least as good as the original strategy. 

In the investigated numerical application, the adaptation of the strategy results only in limited cost reductions in the order of $5-10\%$. The likely reason for this is that the inspection outcomes are not surprising under the prior model. Were one to consider more unexpected inspection outcomes, the adapted optimal heuristic strategy would differ more significantly from the original strategy. 

We introduce a heuristic for prioritizing components for inspections, which is motivated by the value-of-information concept. It accounts for the importance of component in the structural system, but also for the amount of information obtained on other components in the system. Since it is not know a-priori which of these two effects if more important, a heuristic parameter is optimized to weight them. 
In the future, the heuristic might be adjusted in cases when components are not equi-correlated. Thereby it must be considered that inspections on components with higher correlation are likely to provide more information on the system overall. Simultaneously, it should also be ensured that components are selected for inspections in way that ensures their representativeness for the entire structure.
An adjusted heuristic can also account for varying costs of inspections. 

In general, the approach is flexible with respect to adding new heuristic rules and parameters. While there is no guarantee that any given heuristic is optimal, the computed expected total life-cycle cost allows to select the best heuristic among the investigated ones. Hence the more heuristics are investigated, the better. 

We demonstrate the proposed approach by application to a benchmark problem. There are still challenges to its application in practice. Firstly, setting up the probabilistic model of the structure and the deterioration is non-trivial in many cases. Secondly, the translation of the model to a DBN requires careful calibration of the DBN parameters. The first challenge is shared by all approaches that aim at using physics-based models for predictive maintenance. The second challenge might be addressed by writing corresponding software tools that automatize this task. Alternatively, the DBN could also be replaced by another method that enables fast and efficient Bayesian analysis and reliability updating at the structural system level. 

Finally, an extension of the framework to include monitoring data seems straightforward, as long as a Bayesian analysis of the structural system with the monitoring data is possible. Monitoring data can then be treated in the same way as inspection data. With such an extension, the framework can be utilized to optimize monitoring systems.

\section*{Acknowledgements}
We thank Jorge Mendoza for his valuable comments on an earlier
version of this manuscript. This work is supported by the Deutsche Forschungsgemeinschaft (DFG) through Grant STR 1140/3-2 and through the TUM International Graduate School of Science and Engineering (IGSSE).

\appendix
\section{Parameters of the deterioration model}
\FloatBarrier
\setcounter{table}{0}
\setcounter{figure}{0}
\subsection{General model parameters}\label{App:ModelParam}

The parameters of deterioration model are obtained from \citep{Straub_09,Luque_Straub_16}. The relationship between $\text{ln}(C)$ and $M$ is defined after \citep{Ditlevsen_Madsen_96}, such that the joint distribution of $(\text{ln}(C),M)$ is normally distributed with mean value $(-33,3.5)$, standard deviations $(0.47,0.3)$. Furthermore, we assume a linear relationship between $\text{ln}(C)$ and $M$, implying a correlation coefficient of $-1$, in line with \citep{McCartney_Irving_77, Straub_04}.

A time step $t$ corresponds to $\nu=10^5$ fatigue cycles.

For each fatigue hotspots, the stress scale factor $K$ is assumed lognormally distributed. The standard deviation of $\text{ln}(K)$ is $0.22$ for all hotspots, after \citep{Moan_Song_00}. The mean of $\text{ln}(K)$ varies with the assumed fatigue life for each component (see  \ref{App:CalibrationFatigue}).

\begin{table}[ht!]
	\centering
	\caption{Parameters of the fatigue crack growth model. \label{Tab:lDeter}}
	\begin{tabular}{c c c c}
		\hline
		Variable & Type & Mean & Std. Deviation\\
		\hline
		$\alpha_{a_0}$ & Normal dist. & 0 & 1\\
		$\alpha_K$ & Normal dist.& 0 & 1\\
		$\alpha_M$ & Normal dist.& 0 & 1\\
		$D_{k,0}$ [mm] & Exponential dist.& 1 & 1\\
		$M_{k}$  & Normal dist.& 3.5 & 0.3\\
		$\text{ln}K_{k}$  & Normal dist.& Obtained from Table \ref{Tab:K_FDF} & 0.22\\
		$\text{ln}(C_{k,i)}$ & Function & $\text{ln}(C_{k,i})=-1.5667*M_{k,i}-27.5166$\\
		$\lambda$ & Deterministic & $0.8$\\
		$d_{cr}$ [mm] & Deterministic & $50$\\
		$\xi$ [mm] & Deterministic & $10$\\
		$\nu$ [cycles]& Deterministic & $10^5$\\
		$T$ [years]& Deterministic & $40$\\
		$\rho_{D_0}$& Deterministic & $0.5$\\
		$\rho_{M}$& Deterministic & $0.6$\\
		$\rho_{K}$& Deterministic & $0.8$\\
		\hline
	\end{tabular}
\end{table}

\subsection{Calibration of the fatigue stress range parameter to the fatigue life}\label{App:CalibrationFatigue}

Different approaches for the calibration of deterioration models to the component fatigue design factor (FDF) are discussed in detail in \citep{Straub_04}. The FDF is defined as the ratio between the component fatigue life, $T_{FL}$, and the system service life, which here is  $T_{SL}=T=40$[years]. We calibrate the mean value of the random stress scale parameter $K$ in Equation~\ref{Eq:Weibull} to the FDF in two steps, so that two damage models coincide in terms of probability of component failure after $T_{SL}$ years. 

The first model is the Palmgren-Miner damage accumulation law \citep{Palmgren_24,Miner_45}. 
For high-cycle fatigue with $n$ stress cycles, the total accumulated damage $\delta_{n}$ is approximated by
	\begin{equation}\label{Eq:Palmgren}
	\delta_{n}=n\cdot \mathbf{E}_S\left[\frac{1}{N_F(S)}\right],
	\end{equation} 
where $N_F$ is the number of cycles to failure at a constant stress amplitude $S$. 
This relationship, better known as S-N curve, has been defined empirically for various materials, geometries and conditions \citep{Gurney_76,Hobbacher_15}. We adopt the design S-N curve D from the Department of Energy (DoE), UK, after \citep{Straub_04}. It is assumed that the fatigue stress $S$ has a Weibull distribution with scale parameter $k_{S}$ and shape parameter $\lambda_{S}=0.8$.

Failure occurs when the total damage exceeds a critical threshold, $\Delta$, lognormally distributed with moments $(1,0.3)$ \citep{Faber_et_al_00, JCSS}. The resulting limit state function after $n$ cycles is written:
\begin{equation}\label{Eq:LSF_SN}
g_{SN}(n)=\Delta-n \cdot\mathbf{E}_S\left[\frac{1}{N_F(S)}\right].
\end{equation}

Additionally, the fatigue life $FDF\cdot T$ is defined such that the total damage  reaches the critical value 1, with a yearly cycle rate $\nu$, i.e. $n=\nu \cdot FDF \cdot T$:
\begin{equation}\label{Eq:Palmgren_2}
\nu \cdot FDF \cdot T \cdot\mathbf{E}_S\left[\frac{1}{N_F(S)}\right]=1.
\end{equation}

\begin{figure}[t!]
		\centering
			\captionsetup[subfigure]{justification=centering}
	\begin{subfigure}[t]{0.3\columnwidth}
	\includegraphics[trim=55 0 200 0, clip, scale=0.5]{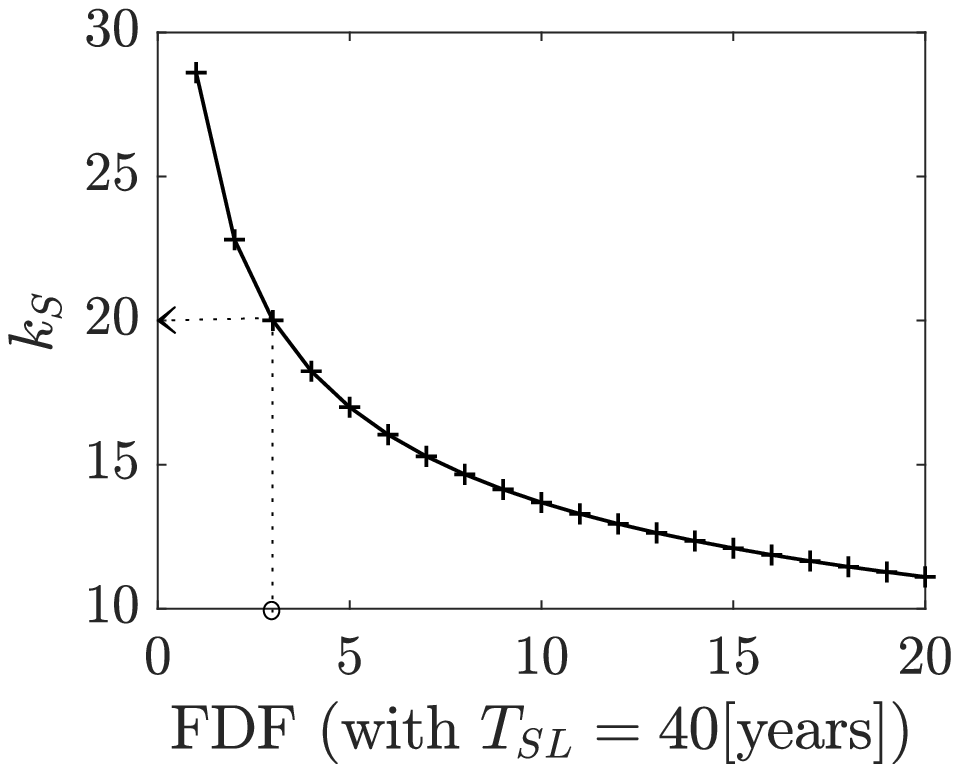}
	\caption{~}
	\label{Fig:FDF_kDS}
\end{subfigure}~
\begin{subfigure}[t]{0.35\columnwidth}
	\includegraphics[trim=15 0 30 0, clip, scale=0.5]{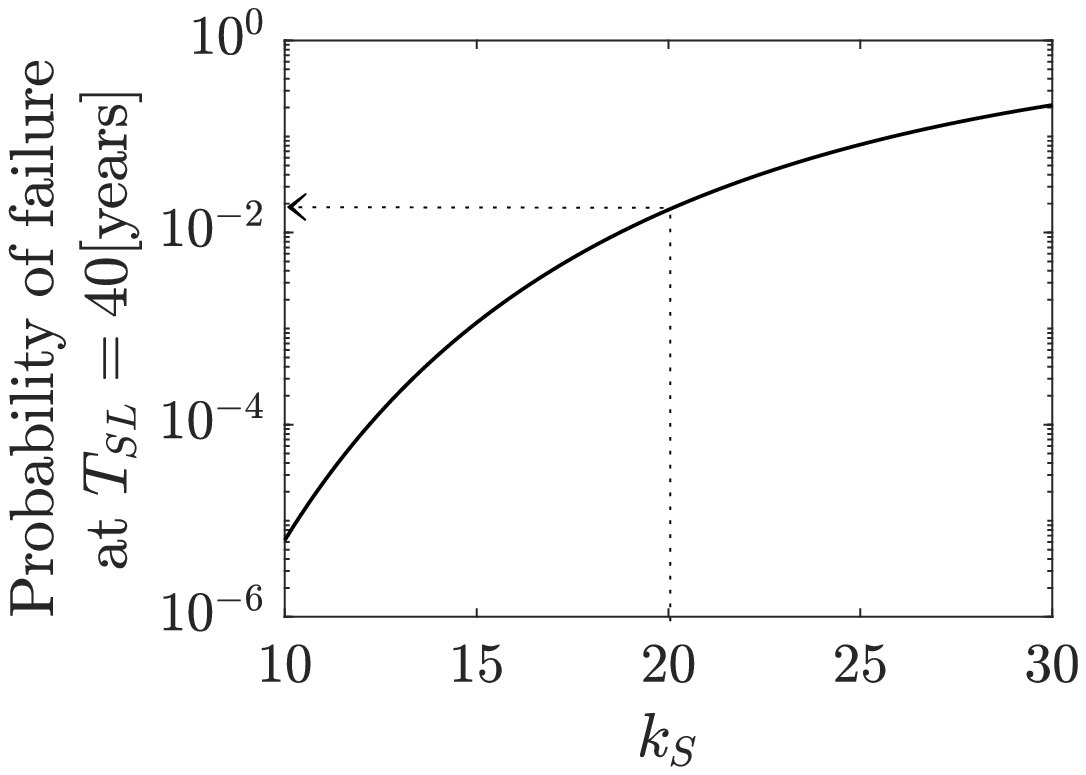}
	\caption{~}
	\label{Fig:kDS_PoF}
\end{subfigure}%
\begin{subfigure}[t]{0.35\columnwidth}
	\includegraphics[trim=15 0 0 0, clip, scale=0.5]{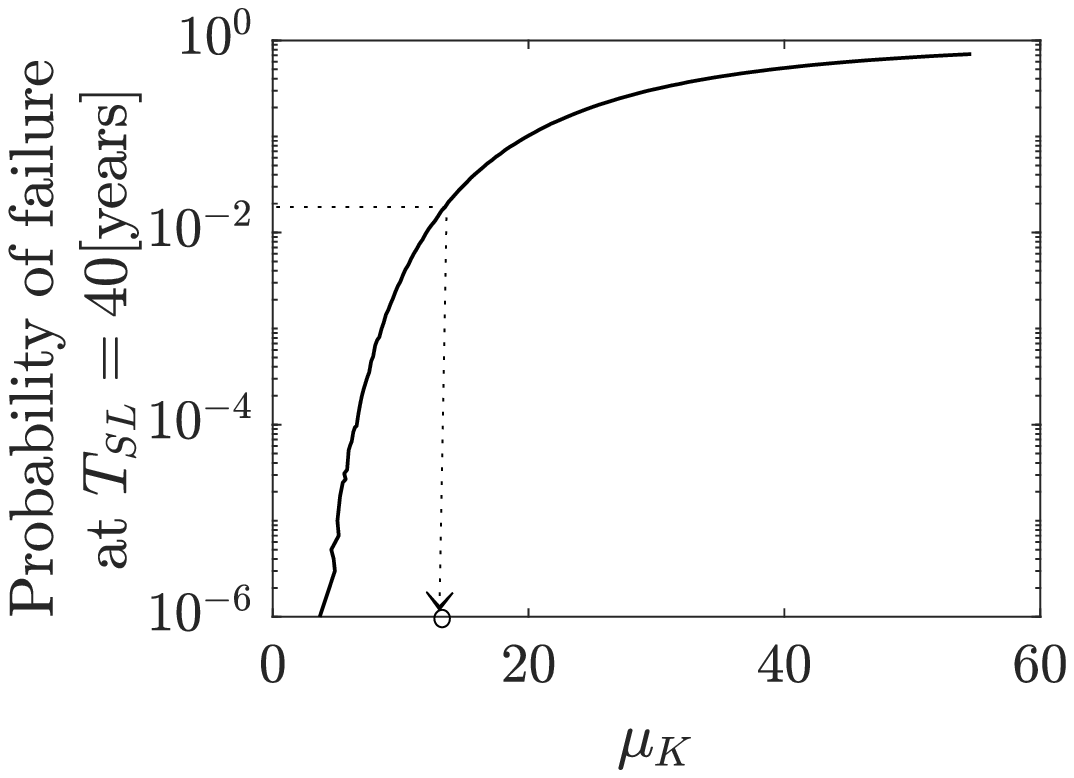}
	\caption{~}
	\label{Fig:meanK_PoF}
\end{subfigure}
\caption{(a): S-N model parameter $k_S$ in function of the FDF. (b): Probability of hotspot failure calculated from the S-N model with parameter $k_S$. (c): Probability of hotspot failure according to the FM model, in function of $\mu_K$. }
\label{Fig:Calibrate}
\end{figure}

The expectation in Equation~\ref{Eq:Palmgren_2} %can be expressed in function $k_{S}$, $\lambda_S$, $m_1$, $m_2$, $C_1^D$ and $N_q$, 
establishes a mapping between FDF and parameter $k_{S}$, depicted in Figure~\ref{Fig:FDF_kDS}. 
Furthermore, for a fixed value of $k_{S}$, the probability of failure at the end of the lifetime $T$,  $\text{Pr}\left[g_{SN}(n=\nu\cdot T)<0\right]$, given by the Palmgren-Miner is approximated with FORM (Figure~\ref{Fig:kDS_PoF}). We note that the evaluation of the probability of failure uses the stochastic description of the S-N curve, while the expectation in Equation~\ref{Eq:Palmgren_2} is computed with the associated characteristic S-N curve.

The second model is the fracture mechanics (FM) model of Equation~\ref{Eq:Paris}. The limit state function defining failure at time $T$ is written as 
\begin{equation}\label{Eq:LSF_FM}
g_{FM}=d_{cr}-D_T,
\end{equation}
where $D_T$ is the crack depth at time $T$, obtained by integration of Equation~\ref{Eq:Paris}. 
It is expressed in function of the random variables $K$, $M$ and $D_0$ (see Section~\ref{Section:Model_Descr}).
The probability of failure $\text{Pr}\left[g_{FM}<0\right]$ in function of the assumed mean value of $K$, $\mu_K$, is shown in Figure~\ref{Fig:meanK_PoF}.
	
Finally, the chosen FDF is mapped to $\mu_K$ such that the limit state functions in Equations~\ref{Eq:LSF_FM} and~\ref{Eq:LSF_SN} result in the same probability of failure at the end of service life $T$. The FDF values and corresponding means of $K$ for all hotspots for the numerical application are obtained by jointly reading Figures~\ref{Fig:Calibrate} (a-c) and are summarized in Table~\ref{Tab:K_FDF}.

\begin{table}[htbp]
	\centering
	\caption{Fatigue design factors (FDFs) and corresponding mean of random variable $K$ at each hotspot. \label{Tab:K_FDF}}
	\begin{tabular}{c c c c}
		\hline
		Hotspot index $k$ & FDF & Mean of $K_{k}$ \\
		\hline
		$\{1,2,3,4\}$ & 10 & $7.58$ \\
		$\{5,6,13,14,17,18,21,22\}$ & 2 & $16.26$ \\
		$\{7,8,9,10,11,12,19,20\}$ & 3 & $13.29$\\
		$\{15,16\}$ & 7 & $8.88$\\
		\hline
	\end{tabular}
\end{table}
\FloatBarrier

\section{Estimation of the annual risk of failure}\label{App:Variance}
The efficiency of the optimization method described in Section~\ref{Section:DPS} depends on the sampling uncertainty of the expectation with MCS (Equation~\ref{Eq:MCS}), with $n_{MC}=1$ as the limiting case. 

In theory, Equation~\ref{Eq:Exp_Risk} coupled with Equation~\ref{Eq:MCS}, implies that the probabilities at every time step should be conditioned on the full observation history $\bm{Z}_{1:n_T}$, which includes the future inspection outcomes. In this section, we show that one can replace the smoothed probability $\text{Pr}(F_i|\bm{w},\bm{Z}_{1:{n_T}})$ in Equation~\ref{Eq:Exp_Risk} by the filtered probability $\text{Pr}(F_i|\bm{w},\bm{Z}_{1:{i-1}})$ when estimating the expected risk of failure. Furthermore, we show that the corresponding Monte Carlo estimator of the risk has a smaller sample variance.

We consider the actions that affect the system reliability $\bm{A}=\{A_1...A_{n_T}\}$ during the service life, such as repair or other maintenance actions. In this paper, the actions are deterministic for given inspection history $\bm{Z}_{1:{n_T}}$ and strategy $\mathcal{S}_{\bm{w}}$,  i.e. $\bm{A}=\bm{A}(\bm{w},\bm{Z}_{1:{n_T}})$. Here we explicitly include them in the expression of the expected cumulative probability of failure:

\begin{equation}\label{Eq:App:1}
\mathbf{E}_{\bm{Z}_{1:{n_T}}}\left[\text{Pr}(F_i|\bm{w},\bm{Z}_{1:{n_T}})\right]=\mathbf{E}_{\bm{Z}_{1:{n_T}}}\left[\text{Pr}(F_i|\bm{A}(\bm{w},\bm{Z}_{1:{n_T}}),\bm{w},\bm{Z}_{1:{n_T}})\right]
\end{equation}

We recall two key principles in a sequential decision process. Firstly, a policy at time step $i$ that assigns an action $A_{i}$ can only consider information $\bm{Z}_{1:{i}}$ about the system up to that time, or in other terms, decisions cannot be based on specific knowledge acquired in the future, i.e. $A_{i}=A_{i}(\bm{w},\bm{Z}_{1:{i}})$. Secondly, the cumulative failure event $F_i$ never depends on the actions after time step $i$, $\bm{A}_{i:n_T}$, neither unconditionally nor conditionally on $\bm{Z}_{1:{n_T}}$. Hence we obtain Equation~\ref{Eq:App:2}:
\begin{equation}\label{Eq:App:2}
\begin{split}
\mathbf{E}_{\bm{Z}_{1:{n_T}}}\left[\text{Pr}(F_i|\bm{w},\bm{Z}_{1:{n_T}})\right]&=\mathbf{E}_{\bm{Z}_{1:{n_T}}}\left[\text{Pr}(F_i|\bm{A}_{1:{i-1}}(\bm{w},\bm{Z}_{1:{i-1}}),\bm{w},\bm{Z}_{1:{n_T}})\right]
\end{split}
\end{equation}

From there, we can split the expectation of the right hand term of Equation~\ref{Eq:App:2}:
\begin{equation}\label{Eq:App:3}
\begin{split}
\mathbf{E}_{\bm{Z}_{1:{n_T}}}\left[\text{Pr}(F_i|\bm{w},\bm{Z}_{1:{n_T}})\right]&=\mathbf{E}_{\bm{Z}_{1:{i-1}}}\left[\mathbf{E}_{\bm{Z}_{{i}:n_T}|{\bm{Z}_{1:{i-1}}}}\left[\text{Pr}(F_i|\bm{A}_{1:{i-1}}(\bm{w},\bm{Z}_{1:{i-1}}),\bm{w},\bm{Z}_{1:{i-1}},\bm{Z}_{{i}:n_T})\right]\right]\\&=\mathbf{E}_{\bm{Z}_{1:{i-1}}}\left[\text{Pr}(F_i|\bm{A}_{1:{i-1}}(\bm{w},\bm{Z}_{1:{i-1}}),\bm{w},\bm{Z}_{1:{i-1}})\right].
\end{split}
\end{equation}

Finally we obtain,
\begin{equation}\label{Eq:App:4}
\begin{split}
\mathbf{E}_{\bm{Z}_{1:{n_T}}}\left[\text{Pr}(F_i|\bm{w},\bm{Z}_{1:{n_T}})\right]&=\mathbf{E}_{\bm{Z}_{1:{i-1}}}\left[\text{Pr}(F_i|\bm{w},\bm{Z}_{1:{i-1}})\right].
\end{split}
\end{equation}

Therefore for a sample observation history $\mathbf{z}_{1:{n_T}}$, the conditional filtered probability, $\text{Pr}(F_i|\bm{w},\mathbf{z}_{1:{i-1}})$, and the smoothed probability, $\text{Pr}(F_i|\bm{w},\mathbf{z}_{1:{n_T}})$, are two valid unbiased estimators of the expected value $\mathbf{E}_{\bm{Z}_{1:{n_T}}}\left[\text{Pr}(F_i|\bm{w},\bm{Z}_{1:{n_T}})\right]$. Furthermore, Equation~\ref{Eq:App:3} shows that the filtered probability is equal to the expectation of the smoothed probability over the inspection results $\bm{Z}_{i:n_T}$. This implies that the filtered probability is an estimator with a smaller variance than the smoothed probability, hence is better suited as estimator of the noisy objective function in the optimization method described in Section~\ref{Section:DPS}.

\begin{figure}[t!]
    \centering
    \captionsetup[subfigure]{justification=centering}
	\begin{subfigure}[t]{0.5\columnwidth}
	\includegraphics[trim=0 0 25 0,clip, scale=0.45]{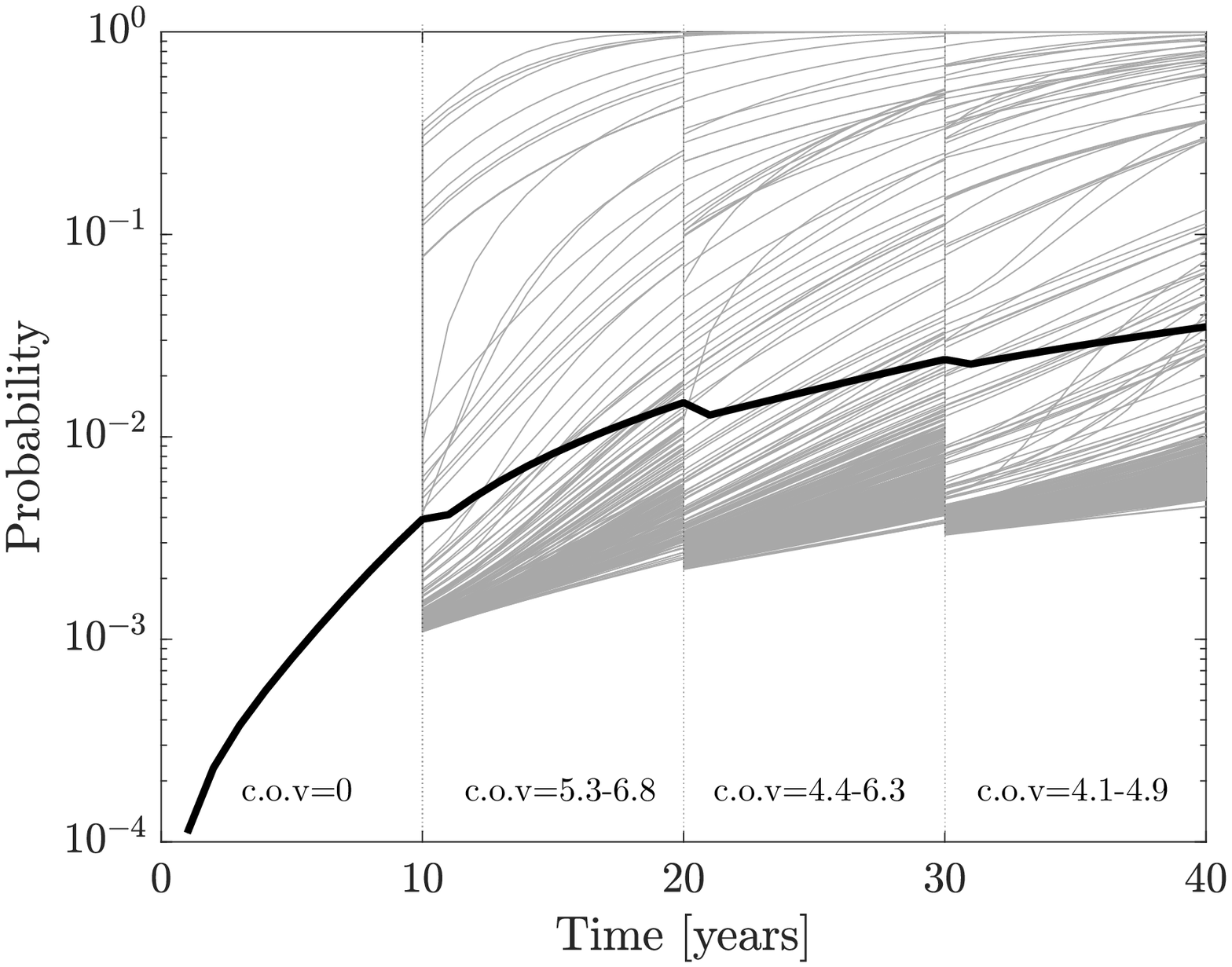}
	\caption{~}\label{Fig:Filtered}
	\end{subfigure}%
	\begin{subfigure}[t]{0.5\columnwidth}
		\includegraphics[trim=0 0 25 0,clip, scale=0.45]{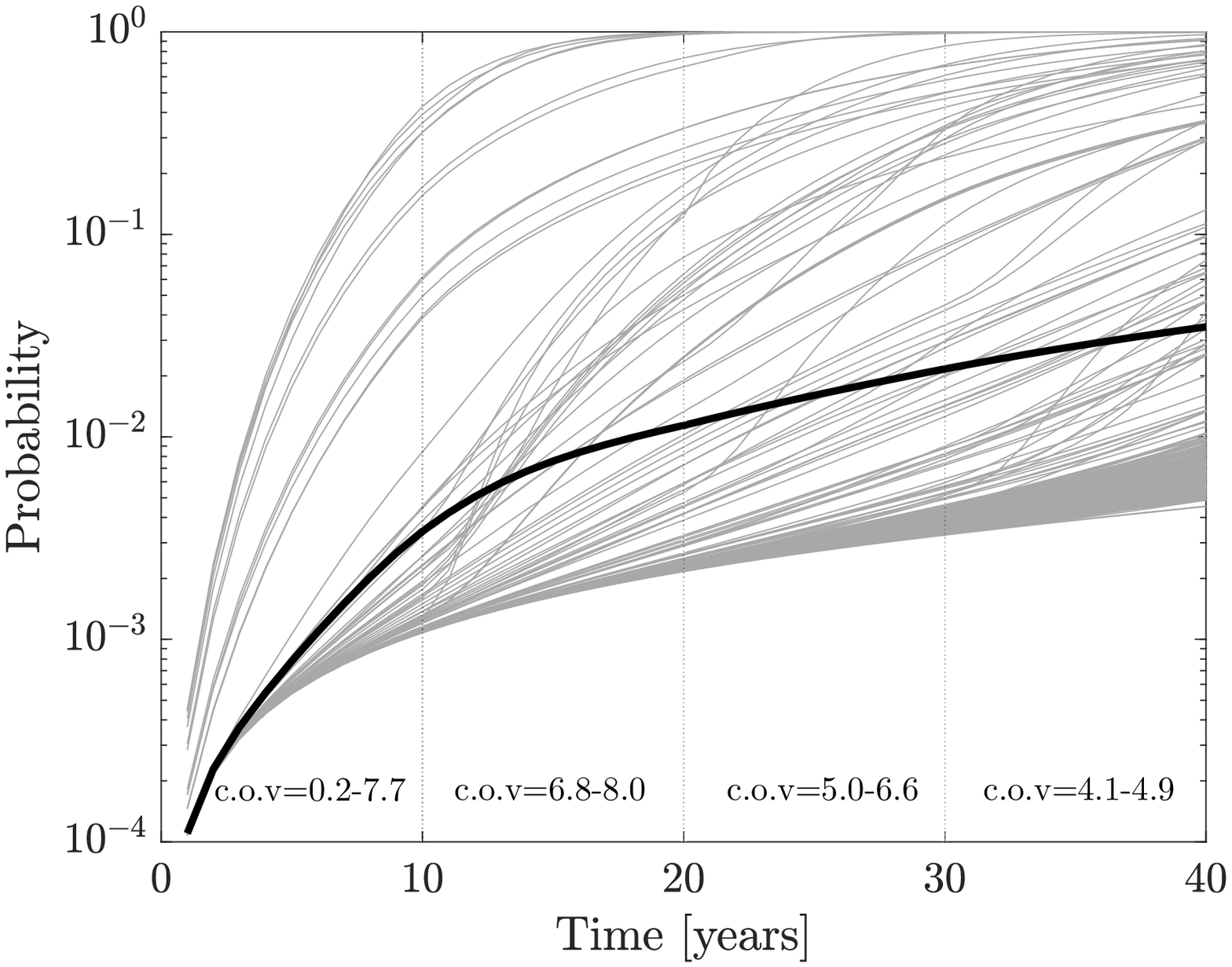}
		\caption{~}\label{Fig:Smoothed}
\end{subfigure}
    \caption{Strategy defined by $\bm{w}=\{ \Delta T=10, p_{th}=1, n_i=5, \eta=1, D_{rep}=\infty\}$. (a): Filtered conditional probability of failure $\text{Pr}(F_i|\bm{w},\bm{Z}_{1:{i-1}})$ for $500$ sample histories $\bm{z}^{(q)}$, and resulting mean (bold line); (b) Smoothed conditional probability of failure $\text{Pr}(F_i|\bm{w},\bm{Z}_{1:n_T})$ for $500$ sample histories $\bm{z}^{(q)}$, and resulting mean (bold line). The estimated range of the coefficient of variation (c.o.v) is indicated for each decade.}
    \label{Fig:Check_Variance}
\end{figure}

The variances of these two estimator and the standard error of the mean can be compared for the I\&M strategies that prescribe only inspections but no repair or maintenance actions to be carried out, i.e. for any $\bm{Z}_{1:{n_T}}$, and at any time step $i$, $A_{i}(\bm{Z}_{1:i})=\textit{`do nothing'}$. For instance, the strategy parametrized by $\bm{w}=\{\Delta T=10, p_{th}=1, n_i=5, \eta=1, D_{rep}=\infty\}$ satisfies this condition. The expectations $\mathbf{E}_{\bm{Z}_{1:{i-1}}}\left[\text{Pr}(F_i|\bm{w},\bm{Z}_{1:{i-1}})\right]$ and $\mathbf{E}_{\bm{Z}_{1:{n_T}}}\left[\text{Pr}(F_i|\bm{w},\bm{Z}_{1:{n_T}})\right]$ are equal to $\text{Pr}(F_i)$ calculated with the prior model assumptions. The variance of these estimators is evaluated for the numerical application described in Section~\ref{Section:Numerical}, with $n_{MC}=500$ sample histories (see Figure~\ref{Fig:Check_Variance}).

The variance of the smoothed estimator is indeed larger than of the filtered estimator: for the times between $1$ and $10$ years, the filtered estimator is exact, and the coefficient of variation of the smoothed estimator is of the order of 500\%. For the subsequent time steps, up to year $30$, the  coefficient of variation of the smoothed estimator is up to 1.3 times that of the filtered estimator.

\section{Prioritization Index for component inspection}\label{App:PI}
The Prioritization Index (PI) is chosen as a proxy for the VoI. The reasoning behind the expression for the PI in Equation~\ref{Eq:PI} is outlined in this section. Here $F_s$ is used to denote the event $F_i^*$ for a given time step $i$. The following derivations omit the conditioning on $\bm{Z}_{1:{i-1}}$. The $SEI_k$ are nonetheless independent of any observation, by definition.

We express the certain event as $\{\Omega\}=\{\{F_{c_{1}}\cup\overline{F_{c_{1}}}\}\cap...\cap\{F_{c_{N}}\cup\overline{F_{c_{N}}}\}\}$, and obtain that
\begin{equation} \label{VoI1}
\begin{split}
\text{Pr}(F_s)  =~ &\text{Pr}(F_s\cap\{\{F_{c_{1}}\cup\overline{F_{c_{1}}}\}\cap...\cap\{F_{c_{N}}\cup\overline{F_{c_{N}}}\}\})\\
=~&\text{Pr}(F_s,\overline{F_{c_{1}}},...,\overline{F_{c_{N}}})+ \text{Pr}(F_s,F_{c_{1}},\overline{F_{c_{2}}},...,\overline{F_{c_{N}}})+\text{Pr}(F_s,\overline{F_{c_{1}}},F_{c_{2}},...\overline{F_{c_{N}}})\\
&+...+\text{Pr}(F_s,\overline{F_{c_{1}}},\overline{F_{c_{2}}},...,\overline{F_{c_{N-1}}},F_{c_{N}})+ {\sum_{\mathcal{I}}{\text{Pr}(F_s,{F_{c_{\mathcal{I}}}}\cap{\overline{F_{c_{\smallsetminus \mathcal{I}}}}})}}\\
\end{split}
\end{equation}
The last term in Equation~\ref{VoI1} corresponds to the joint probabilities of $F_s$ and failed components belonging to the subsets $\mathcal{I}$ of $\{1..N\}$, with $|\mathcal{I}|\geq 2$.
%\sum_{k=2}^{N}{\sum_{\mathscr{P}_k}{\text{Pr}(F_s,\bigcap_{i \in{\mathscr{P}_k}}{F_{c_{i}}},\bigcap_{j \in{\overline{\mathscr{P}_k}}}{\overline{F_{c_{j}}}})}}\\

By introducing conditional probabilities and writing $a=\text{Pr}(F_s|\overline{F_{c_{1}}},...,\overline{F_{c_{N}}})$, the probability of failure of the system is
\begin{equation} \label{Eq:VoI1_final}
\begin{split}
\text{Pr}(F_s)  =&~a\cdot \text{Pr}(\overline{F_{c_{1}}},...,\overline{F_{c_{N}}})+ \text{Pr}(F_s|F_{c_{1}},\overline{F_{c_{2}}},...,\overline{F_{c_{N}}})\cdot \text{Pr}(F_{c_{1}},\overline{F_{c_{2}}},...,\overline{F_{c_{N}}})+...+\\
&\text{Pr}(F_s|\overline{F_{c_{1}}},\overline{F_{c_{2}}},...F_{c_{N}})\cdot \text{Pr}(\overline{F_{c_{1}}},\overline{F_{c_{2}}},...F_{c_{N}})+{\sum_{\mathcal{I}}{\text{Pr}(F_s|{F_{c_{\mathcal{I}}}}\cap{\overline{F_{c_{\smallsetminus \mathcal{I}}}}})\cdot \text{Pr}({F_{c_{\mathcal{I}}}}\cap{\overline{F_{c_{\smallsetminus \mathcal{I}}}}})}}\\
\end{split}
\end{equation}
From Equation~\ref{Eq:SEI}, we have that $\text{Pr}(F_{s}|\overline{F_{c_{1}}},...,\overline{F_{c_{k-1}}},F_{c_k},\overline{F_{c_{k+1}}},...,\overline{F_{c_{N}}})=a+SEI_k$, and similarly we can express each $\text{Pr}(F_s|{F_{c_{\mathcal{I}}}}\cap{\overline{F_{c_{\smallsetminus \mathcal{I}}}}})=a+MEI_\mathcal{I}$, where $MEI_\mathcal{I}$ is the multiple elements importance of components in $\mathcal{I}$. By factorizing $a$, we obtain Equation~\ref{Eq:VoI2}:
\begin{equation} \label{Eq:VoI2}
\begin{split}
\text{Pr}(F_s)  =&~a+ SEI_1\cdot \text{Pr}(F_{c_{1}},\overline{F_{c_{2}}},...,\overline{F_{c_{N}}})+...\\
&+SEI_N\cdot \text{Pr}(\overline{F_{c_{1}}},\overline{F_{c_{2}}},...F_{c_{N}})+b,
\end{split}
\end{equation}
where $b$ represents the contribution of simultaneous component failures.  Equation~\ref{Eq:VoI3} introduces the marginal probabilities of component failure:
\begin{equation} \label{Eq:VoI3}
\begin{split}
\text{Pr}(F_s)  =&~a+ SEI_1\cdot \text{Pr}(F_{c_{1}}) \cdot \text{Pr}(\overline{F_{c_{2}}},...,\overline{F_{c_{N}}}|F_{c_{1}})\\
&+...+SEI_N \cdot \text{Pr}(F_{c_{N}})\cdot \text{Pr}(\overline{F_{c_{1}}},...,\overline{F_{c_{N-1}}}|F_{c_{N}})+b.
\end{split}
\end{equation}

From Equation~\ref{Eq:VoI3}, we approximate $\text{Pr}(F_s)$ with
\begin{equation} \label{Eq:VoI3_final}
\text{Pr}(F_s) \simeq  a + SEI_1\cdot \text{Pr}(F_{c_{1}})+...+SEI_N \cdot \text{Pr}(F_{c_{N}}) + b,
\end{equation}

 Equation~\ref{Eq:VoI3_final} is not actually used to calculate the interval failure probability of the system, $\text{Pr}(F_i^*)$, but shows that the probability of system failure approximately is a function of the terms $SEI_k\cdot \text{Pr}(F_{c_{k}})$ defined for each component, hence by the VoI for component $k$ is a linear function of $SEI_k\cdot \text{Pr}(F_{c_{k}})$. Furthermore, the amount of information learnt on other components is related to the probability of failure $\text{Pr}(F_{c_{k}})$, through the components’ interdependence.

\begin{figure}[t!]
	\centering
	\captionsetup[subfigure]{justification=centering}
	
	\begin{subfigure}[t]{0.5\linewidth}
		
		\includegraphics[trim=30 0 30 0, width=0.95\linewidth]{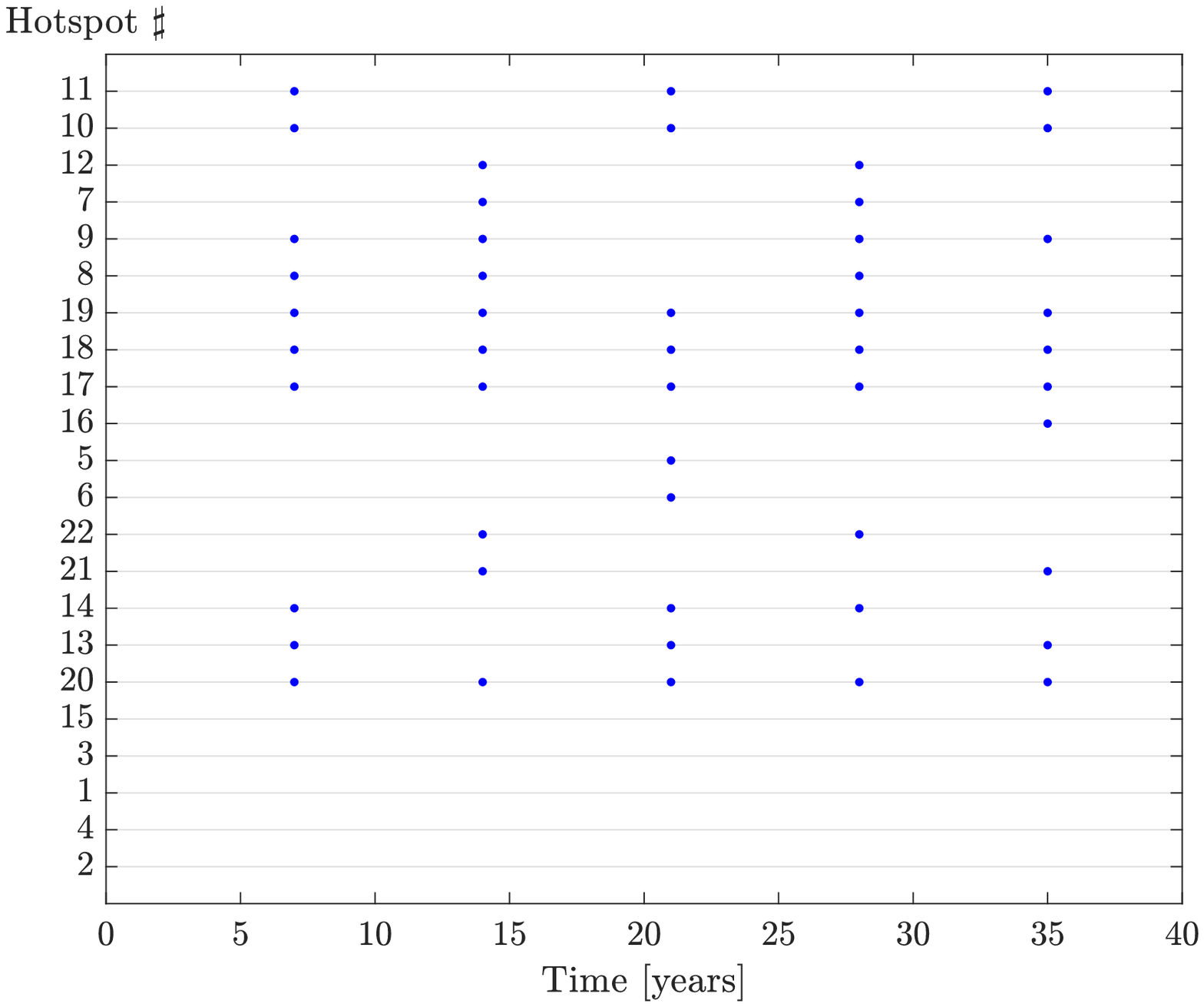}
		\caption{$\eta=0.84$}
		
	\end{subfigure}%
	\begin{subfigure}[t]{0.5\linewidth}
		
		\includegraphics[trim=30 0 30 0,width=0.95\linewidth]{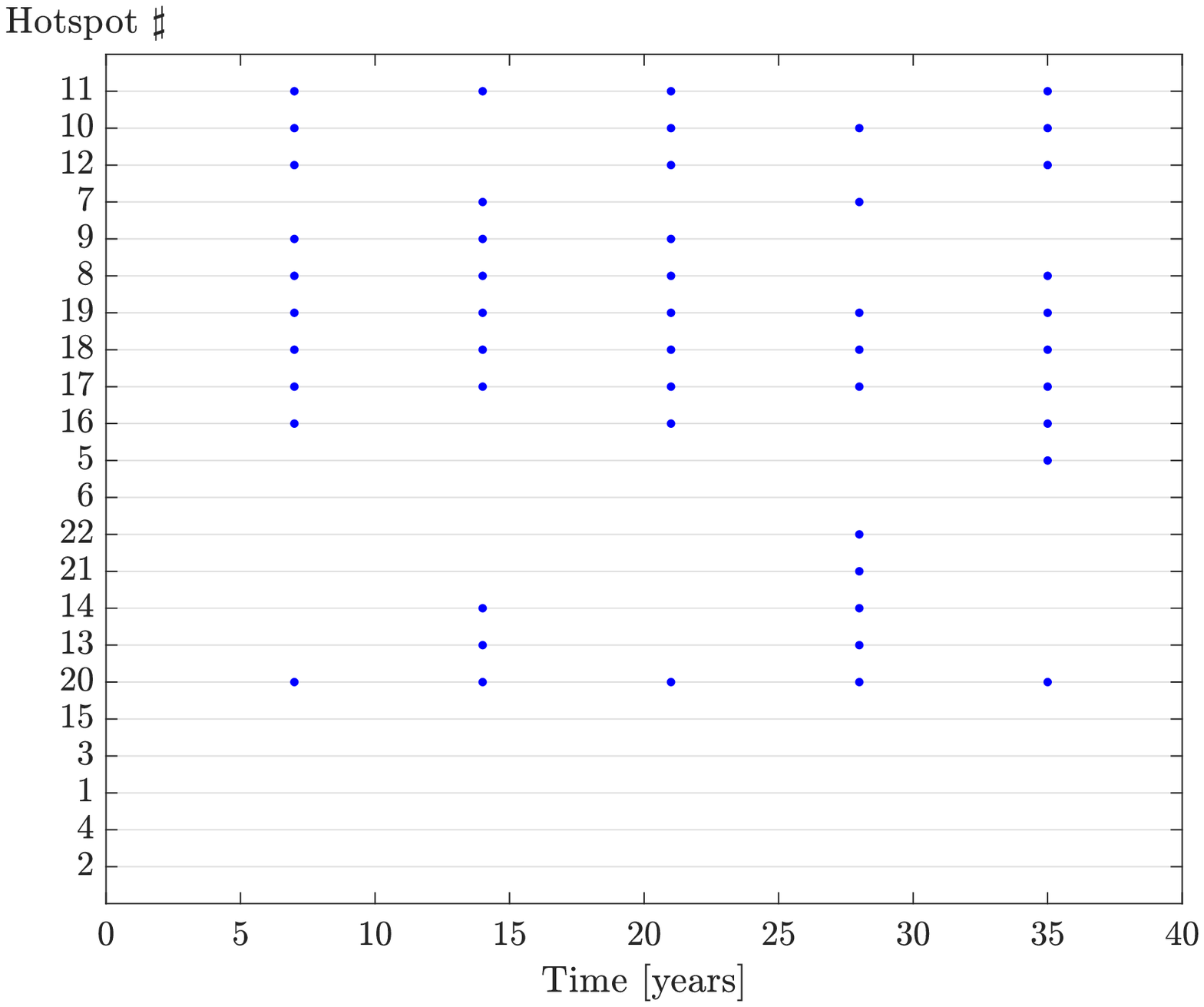}
		\caption{$\eta=2.96$}
		
	\end{subfigure}
	\caption{History of hotspots inspected (marked with a blue dot) for a sample deterioration history, following the strategies defined by $\Delta T=7$[years], $n_I=10$ and $p_{th}=5.2\cdot10^{-3}$, and a lower (a) and higher (b) prioritization parameter $\eta$. The hotspots numbers on the y-axis are sorted according to their calculated $SEI$.}
	\label{Fig:Compare_eta}
\end{figure}

Figure~\ref{Fig:Compare_eta} compares two sample inspection histories for two values of $\eta$, from the numerical application to the Zayas frame. As expected, more hotspots are inspected during the service life when a lower value for $\eta$ is fixed. For a higher value of $\eta$ the inspected hotspots are principally those with the higher $SEI$ values.
\FloatBarrier

%\section*{References}
%\begin{thebibliography}{00}
	
	
	
	\bibliographystyle{elsarticle-harv}
	%\bibliography{Paper_IandM.bib}
	\bibliography{My_EndNote_Library_new}
%		\bibitem{}

%\end{thebibliography}

\end{document}